\pdfoutput=1

\RequirePackage[l2tabu, orthodox]{nag}

\documentclass[11pt,
    final,
    onecolumn,
]{article}

\usepackage[english]{babel}

\usepackage[T1]{fontenc} 
\usepackage[utf8]{inputenc} 
\usepackage{microtype}
\usepackage{lmodern} 
\usepackage{csquotes}

\usepackage{amsmath, amsfonts, amssymb} 
\let\origlangle\langle 
\let\origrangle\rangle 
\usepackage[matha,mathx]{mathabx} 
\let\langle\origlangle 
\let\rangle\origrangle 

\usepackage[affil-it]{authblk} 
\usepackage[useregional]{datetime2} 

\usepackage{graphicx}

\usepackage{booktabs}
\newcommand{\footA}{$^\mathrm{a}$}
\newcommand{\footB}{$^\mathrm{b}$}

\usepackage[
    backend = bibtex,
    uniquename = init,
    giveninits = true,
    sortcites = true,
    sorting = nyt,
    sortlocale = en_US,
    style = numeric, 
    maxbibnames = 99,
]{biblatex}
\DeclareNameAlias{sortname}{last-first} 
\DeclareNameAlias{default}{last-first} 

\AtEveryBibitem{\clearfield{month}}
\AtEveryBibitem{\clearfield{day}}

\AtEveryCitekey{\iffieldsequal{eid}{pages}{\clearfield{pages}}{}}
\AtEveryBibitem{\iffieldsequal{eid}{pages}{\clearfield{pages}}{}}

\ifpdf
\usepackage[
    bookmarks=true,
    bookmarksnumbered=true,
    bookmarksopen=true,
    colorlinks=true,
    allcolors=blue,
    pdfborder={0 0 0}
]{hyperref}
\fi

\usepackage[capitalize, sort]{cleveref}

\ifpdf
    \hypersetup{pdftitle={A new subgrid characteristic length for turbulence simulations on anisotropic grids},pdfauthor={F. X. Trias, A. Gorobets, M. H. Silvis, R. W. C. P. Verstappen, A. Oliva}}
\fi

\newcommand{\mydate}{
    Received: \DTMdate{2017-05-12}.
    Revised: \DTMdate{2017-10-16}.
    Accepted: \DTMdate{2017-11-06}. \\
    Published online: \DTMdate{2017-11-29}.
}

\usepackage[totalwidth=480pt, totalheight=680pt]{geometry}

\usepackage{afterpage}

\title{\vspace{-0.6\baselineskip} A new subgrid characteristic length for turbulence simulations on anisotropic grids}
\author[1]{F.~X.~Trias\thanks{\href{mailto:xavi@cttc.upc.edu}{xavi@cttc.upc.edu}}$^,$}
\author[1,2]{A.~Gorobets\thanks{\href{mailto:andrey@cttc.upc.edu}{andrey@cttc.upc.edu}}$^,$}
\author[3]{M.~H.~Silvis\thanks{\href{mailto:m.h.silvis@rug.nl}{m.h.silvis@rug.nl}}$^,$}
\author[3]{R.~W.~C.~P.~Verstappen\thanks{\href{mailto:r.w.c.p.verstappen@rug.nl}{r.w.c.p.verstappen@rug.nl}}$^,$}
\author[1]{A.~Oliva\thanks{\href{mailto:oliva@cttc.upc.edu}{oliva@cttc.upc.edu}}$^,$\vspace{0.8\baselineskip}}
\affil[1]{Heat and Mass Transfer Technological Center, Technical University of Catalonia, C/Colom 11, 08222 Terrassa, Spain}
\affil[2]{Keldysh Institute of Applied Mathematics, 4A, Miusskaya Sq., Moscow 125047, Russia}
\affil[3]{Johann Bernoulli Institute for Mathematics and Computer Science, University of Groningen, P.O. Box 407, 9700 AK Groningen, The Netherlands\vspace{-0.8\baselineskip}}
\date{\mydate}

\providecommand\bcdot{\boldsymbol{\cdot}}

\newcommand{\ud}{\mathrm{d}}

\newcommand\mnewcommand[1]{%
\let#1\relax \newcommand#1 }

\newcommand{\F}[1]{\overline{#1}}

\newcommand{\ie}{\textit{i.e.,}}
\newcommand{\eg}{\textit{e.g.,}}

\newcommand{\etal}{\textit{et al.}}

\newcommand{\kc}{k_{c}}

\mnewcommand{\mysubsubsection}{\subsubsection}
\mnewcommand{\comentar}[1]{}

\mnewcommand{\Nx}{N_\x}
\mnewcommand{\Ny}{N_\y}
\mnewcommand{\Nz}{N_\z}
\mnewcommand{\Nm}{N_m}
\mnewcommand{\Ns}{N_s}
\mnewcommand{\complconj}[1]{#1^{*}}
\mnewcommand{\mvbrack}[1]{\left[ #1 \right]}
\mnewcommand{\step}{\Delta}
\mnewcommand{\dt}{\step t}
\mnewcommand{\traspose}{^{T}}
\mnewcommand{\avgtime}[1]{\left< #1 \right>}
\mnewcommand{\avg}[1]{\overline{#1}}
\mnewcommand{\mcdot}{\bcdot}
\mnewcommand{\mnabla}{\nabla}

\mnewcommand{\real}{\mathbb{R}}
\mnewcommand{\complex}{\mathbb{C}}
\mnewcommand{\imag}{\mathbb{I}}

\mnewcommand{\foutrans}[1]{\hat{#1}}
\mnewcommand{\modetrans}[1]{\tilde{#1}}

\mnewcommand{\timeintparam}{\kappa}

\mnewcommand{\x}{x}
\mnewcommand{\y}{y}
\mnewcommand{\z}{z}

\mnewcommand{\vels}{u_s}
\mnewcommand{\uvel}{u}
\mnewcommand{\vvel}{v}
\mnewcommand{\wvel}{w}

\mnewcommand{\vortx}{\omega_\x}
\mnewcommand{\vorty}{\omega_\y}
\mnewcommand{\vortz}{\omega_\z}

\mnewcommand{\facevel}{[\velh]_\nface}

\mnewcommand{\flux}{f}
\mnewcommand{\Dx}{\Delta x}
\mnewcommand{\Dy}{\Delta \y}
\mnewcommand{\Dz}{\Delta z}

\mnewcommand{\isvector}[1]{\boldsymbol{#1}}
\mnewcommand{\istensor}[1]{\mathsf{#1}}

\mnewcommand{\va}{\isvector{a}}
\mnewcommand{\vb}{\isvector{b}}
\mnewcommand{\vc}{\isvector{c}}
\mnewcommand{\vd}{\isvector{d}}
\mnewcommand{\sca}{\phi}
\mnewcommand{\scafield}{\isvector{\phi}}

\mnewcommand{\vel}{\isvector{u}}
\mnewcommand{\velv}{\isvector{v}}
\mnewcommand{\velw}{\isvector{w}}
\mnewcommand{\normal}{\isvector{n}}

\mnewcommand{\vela}{\vel_{a}}
\mnewcommand{\mua}{\mu_{a}}

\mnewcommand{\velhcsol}{\velhc^{sol}} 

\mnewcommand{\Rh}{R_{h}}

\mnewcommand{\veluc}{\isvector{u}_1}
\mnewcommand{\velvc}{\isvector{u}_2}
\mnewcommand{\velwc}{\isvector{u}_3}

\mnewcommand{\basis}{\isvector{w}}

\mnewcommand{\tensor}{\istensor{T}}
\mnewcommand{\Identity}{\istensor{I}}

\mnewcommand{\bodyforce}{\isvector{f}}
\mnewcommand{\velh}{\vel_s}
\mnewcommand{\velhv}{\velv_s}
\mnewcommand{\velhw}{\velw_s}
\mnewcommand{\tildevelh}{\tilde{\vel}_s}
\mnewcommand{\velhc}{\vel_c}
\mnewcommand{\velhcCB}[1]{\ifthenelse{\equal{#1}{}}{\velhc^{\ominus}}{\velhc^{#1,\ominus}}}
\mnewcommand{\velhcCBFree}[1]{\ifthenelse{\equal{#1}{}}{\velhc^{\oplus}}{\velhc^{#1,\oplus}}}
\mnewcommand{\velhcHF}[1]{\ifthenelse{\equal{#1}{}}{\velhc^{>}}{\velhc^{#1,>}}}
\mnewcommand{\velhcLF}[1]{\ifthenelse{\equal{#1}{}}{\velhc^{<}}{\velhc^{#1,<}}}
\mnewcommand{\velhsHF}[1]{\ifthenelse{\equal{#1}{}}{\velh^{>}}{\velh^{#1,>}}}
\mnewcommand{\velhsLF}[1]{\ifthenelse{\equal{#1}{}}{\velh^{<}}{\velh^{#1,<}}}
\mnewcommand{\vortd}{\vort_v}
\mnewcommand{\presh}{\isvector{p}_c}
\mnewcommand{\bodyforceh}{\bodyforce_c}
\mnewcommand{\diver}{\mnabla \cdot}
\mnewcommand{\lapl}{\mnabla^2}
\mnewcommand{\grad}{\mnabla}
\mnewcommand{\Ru}[1]{\isvector{R} \left( #1 \right)}
\mnewcommand{\Rud}[1]{\mathsfbi{R} \left( #1 \right)}
\mnewcommand{\vecnull}{\isvector{0}}
\mnewcommand{\vecnulls}{\vecnull_{s}}
\mnewcommand{\vecnullg}{\vecnull_{h}}
\mnewcommand{\vecnullc}{\vecnull_{c}}
\mnewcommand{\vecone}{\isvector{1}}
\mnewcommand{\veconec}{\vecone_{c}}
\mnewcommand{\vecones}{\vecone_{s}}
\mnewcommand{\veconeg}{\vecone_{h}}
\mnewcommand{\veconetresc}{\vecone_{3c}}

\mnewcommand{\velhg}{\vel_h}
\mnewcommand{\preshg}{\isvector{p}_h}

\mnewcommand{\dim}{3}  

\mnewcommand{\ud}{d}
\mnewcommand{\vort}{\isvector{\omega}}
\mnewcommand{\rot}[1]{\mnabla \times #1}
\mnewcommand{\selfinnerprod}[1]{\innerprod{#1}{#1}}
\mnewcommand{\innerprod}[2]{( #1 , #2 )}
\mnewcommand{\convective}[2]{C \left( #1 , #2 \right)}
\mnewcommand{\intvol}[1]{\int_{\Omega} #1 \ud \Omega}
\mnewcommand{\intsurf}[1]{\int_{\partial \Omega} #1 \ud S}

\mnewcommand{\nvc}{k}
\mnewcommand{\nedge}{v}
\mnewcommand{\axis}{i}
\mnewcommand{\nface}{f}
\mnewcommand{\cp}{c}
\mnewcommand{\cpA}{{c1}}
\mnewcommand{\cpB}{{c2}}
\mnewcommand{\Fedge}[1]{F_e ( #1 )}
\mnewcommand{\Fcell}[1]{F_f ( #1 )}
\mnewcommand{\Fvolume}[1]{F_c ( #1 )}

\mnewcommand{\mathsfbi}[1]{\mathsf{#1}}

\mnewcommand{\conv}{\mathsfbi{C}\left( \velh \right)}
\mnewcommand{\convg}{\mathsfbi{C}\left( \velhg \right)}
\mnewcommand{\convc}{\mathsfbi{C}_{c} \left( \velh \right)}
\mnewcommand{\convvc}{\mathsfbi{C}_{c}^{\dim d} \left( \velh \right)}
\mnewcommand{\convu}{\mathsfbi{C}_{u} \left( \velh \right)}
\mnewcommand{\convtraspose}{\mathsfbi{C}\traspose\left( \velh \right)}
\mnewcommand{\convgtraspose}{\mathsfbi{C}\traspose\left( \velhg \right)}
\mnewcommand{\convarg}[1]{\mathsfbi{C}\left( #1 \right)}
\mnewcommand{\convargtraspose}[1]{\mathsfbi{C}\traspose\left( #1 \right)}
\mnewcommand{\convutraspose}{\mathsfbi{C}_{u} \traspose\left( \velh \right)}
\mnewcommand{\convmat}{\mathsfbi{C}}

\DeclareRobustCommand{\spreg}[2]{\ifthenelse{\equal{#2}{}}{\convmat_{#1}}{\convmat_{#1}\left( #2 \right)}}
\mnewcommand{\velauxc}{\velv_c}
\mnewcommand{\velauxWc}{\velw_c}

\mnewcommand{\velhcmode}[1]{\ifthenelse{\equal{#1}{+}}{\velhcHF{}}{\ifthenelse{\equal{#1}{-}}{\velhcLF{}}{\ifthenelse{\equal{#1}{=}}{\velhc}{}}}}
\mnewcommand{\velhsmode}[1]{\ifthenelse{\equal{#1}{+}}{\velhsHF{}}{\ifthenelse{\equal{#1}{-}}{\velhsLF{}}{\ifthenelse{\equal{#1}{=}}{\velh}{}}}}

\mnewcommand{\velhctype}[2]{\ifthenelse{\equal{#2}{f}}{\overline{\velhcmode{#1}}}{\ifthenelse{\equal{#2}{r}}{\left(\velhcmode{#1}\right)^\prime}{\ifthenelse{\equal{#2}{v}}{\velhcmode{#1}}{}}}}
\mnewcommand{\velhstype}[2]{\ifthenelse{\equal{#2}{f}}{\overline{\velhsmode{#1}}}{\ifthenelse{\equal{#2}{r}}{\left(\velhsmode{#1}\right)^\prime}{\ifthenelse{\equal{#2}{v}}{\velhsmode{#1}}{}}}}

\mnewcommand{\triadic}[6]{\left(\velhctype{#1}{#2}\right)\traspose \convarg{\velhstype{#3}{#4}} \velhctype{#5}{#6}}

\mnewcommand{\OP}{J}
\mnewcommand{\OPh}{\mathsfbi{\OP}}
\mnewcommand{\OPc}{{\cal \OP}}
\mnewcommand{\diff}{\mathsfbi{D}}
\mnewcommand{\diffg}{\mathsfbi{D}}
\mnewcommand{\diffc}{\diff_{\cp}}
\mnewcommand{\diffvc}{\diffc^{\dim d}}
\mnewcommand{\diffu}{\diff_{u}}
\mnewcommand{\dive}{\mathsfbi{M}}
\mnewcommand{\dives}{\dive_s}
\mnewcommand{\diveg}{\mathsfbi{M}_h}
\mnewcommand{\divescaf}{\dive_{sf}}
\mnewcommand{\divevecf}{\dive_{vf}}
\mnewcommand{\graddc}{\mathsfbi{G}_c}
\mnewcommand{\gradd}{\mathsfbi{G}}
\mnewcommand{\graddg}{\mathsfbi{G}_h}
\mnewcommand{\lapld}{\mathsfbi{L}}
\mnewcommand{\lapldc}{\lapld_{\cp}}
\mnewcommand{\vcvects}{\mathsfbi{\Omega}_s}
\mnewcommand{\pseudovcvects}{\tilde{\mathsfbi{\Omega}}_s}
\mnewcommand{\vcvectg}{\mathsfbi{\Omega}_h}
\mnewcommand{\vcvectc}{\mathsfbi{\Omega}_{\cp}}
\mnewcommand{\vcvectv}{\mathsfbi{\Omega}_v}
\mnewcommand{\tildevcvectv}{\widetilde{\mathsfbi{\Omega}}_v}
\mnewcommand{\tildevcvects}{\widetilde{\mathsfbi{\Omega}}_s}
\mnewcommand{\vcvectvc}{\vcvectc^{\dim d}} 
\mnewcommand{\vcvect}{\mathsfbi{\Omega}}
\mnewcommand{\nullmat}{\mathsfbi{0}}
\mnewcommand{\normd}[1]{|| #1 ||}
\mnewcommand{\rotd}{\mathsfbi{R}}
\mnewcommand{\graddscaf}{\gradd_{sf}}
\mnewcommand{\graddvecf}{\gradd_{vf}}
\mnewcommand{\graddx}{\gradd_{\x}}
\mnewcommand{\graddy}{\gradd_{\y}}
\mnewcommand{\graddz}{\gradd_{\z}}
\mnewcommand{\graddd}[1]{\ifthenelse{\equal{#1}{1}}{\graddx}{\ifthenelse{\equal{#1}{2}}{\graddy}{\ifthenelse{\equal{#1}{3}}{\graddz}{\gradd_{x_i}}}}}
\mnewcommand{\graddprod}[2]{\graddd{#1}\traspose \vcvect \graddd{#2}}

\mnewcommand{\fluxh}[2]{T_{#1}\left({#2}\right)}

\mnewcommand{\twoD}{two-dimensional}
\mnewcommand{\threeD}{three-dimensional}
\mnewcommand{\TwoD}{Two-dimensional}
\mnewcommand{\ThreeD}{Three-dimensional}
\mnewcommand{\biD}{\twoD~}
\mnewcommand{\triD}{\threeD~}
\mnewcommand{\BiD}{\TwoD~}
\mnewcommand{\TriD}{\ThreeD~}
\mnewcommand{\biandtriD}{two- and \threeD~}
\mnewcommand{\BiandtriD}{Two- and \threeD~}
\mnewcommand{\Dim}[1]{\ifthenelse{\equal{#1}{2}}{\twoD}{\ifthenelse{\equal{#1}{3}}{\threeD}{KK}}}

\mnewcommand{\inttypeoftext}{\mathsf}
\mnewcommand{\kinener}{\inttypeoftext{E}}
\mnewcommand{\enstrophy}{\inttypeoftext{\Omega}}
\mnewcommand{\helicity}{\inttypeoftext{H}}
\mnewcommand{\vorthelicity}{\helicity_{\vort}}
\mnewcommand{\palinstrophy}{\inttypeoftext{P}}

\mnewcommand{\helicityd}{\helicity_c}
\mnewcommand{\enstrophyd}{\enstrophy_c}

\mnewcommand{\vvlength}{m}
\mnewcommand{\pvlength}{n}
\mnewcommand{\evlength}{e}

\mnewcommand{\Isc}{\Gamma}
\mnewcommand{\Ics}{\Isc\traspose}

\mnewcommand{\Scs}{\Gamma_{c \rightarrow s}}
\mnewcommand{\Ssc}{\Gamma_{s \rightarrow c}}

\mnewcommand{\Sscal}{\Pi_{c \rightarrow s}}
\mnewcommand{\Sthreescal}{\Pi}
\mnewcommand{\NormalVect}[1]{\ifthenelse{\equal{#1}{}}{\istensor{N}_{s}}{\istensor{N}_{s,#1}}}

\mnewcommand{\Correction}{\istensor{P}}
\mnewcommand{\PseudoCorrection}{\tilde{\Correction}}
\mnewcommand{\kernel}[1]{Ker \left( #1 \right)}

\mnewcommand{\Ivs}{\Psi}
\mnewcommand{\Isv}{\Psi\traspose}

\mnewcommand{\OIsc}{\Upsilon}
\mnewcommand{\OIcs}{\Pi}

\mnewcommand{\order}{o}
\mnewcommand{\coarsemesh}{i}
\mnewcommand{\sizevc}{{V}}
\mnewcommand{\error}{\epsilon}

\mnewcommand{\convH}{\text{(Conv)}_{\helicityd}}
\mnewcommand{\diffH}{\text{(Diff)}_{\helicityd}}
\mnewcommand{\presH}{\text{(Pres)}_{\helicityd}}

\newcommand{\Flength}{\Delta}
\newcommand{\cFlength}{\overline{\Flength}}

\renewcommand{\P}{P}
\newcommand{\Q}{Q}
\newcommand{\R}{R}

\newcommand{\nut}{\nu_e}

\newcommand{\J}{\istensor{J}}  
\newcommand{\G}{\istensor{G}}
\renewcommand{\S}{\istensor{S}}
\renewcommand{\O}{\istensor{\Omega}}
\newcommand{\A}{\istensor{A}}
\newcommand{\DelTen}{\istensor{\Flength}}
\newcommand{\GD}{\G_{\Flength}}

\newcommand{\QA}{\Q_{\A}}

\newcommand{\trace}{\mathrm{tr}}
\newcommand{\tr}[1]{\trace(#1)}
\newcommand{\trpow}[2]{\trace^{#1}(#2)}
\newcommand{\traceless}[1]{{\tilde{#1}}}

\newcommand{\diag}{\mathrm{diag}}
\newcommand{\Det}{\mathrm{det}}
\newcommand{\Circ}{\mathrm{circ}}

\newcommand{\GGt}{\G\G\traspose}
\newcommand{\GtG}{\G\traspose\G}

\newcommand{\KH}{Kelvin--Helmholtz~}
\newcommand{\VK}{von K\'{a}rm\'{a}n~}

\newcommand{\FLvol}{\Flength_{\mathrm{vol}}}
\newcommand{\FLSco}{\Flength_{\mathrm{Sco}}}
\newcommand{\FLmax}{\Flength_{\max}}
\newcommand{\FLLtwo}{\Flength_{\mathrm{L2}}}
\newcommand{\FLLapl}{\Flength_{\mathrm{Lapl}}}
\newcommand{\FLvort}{\Flength_{\vort}}
\newcommand{\FLMoc}{\tilde{\Flength}_{\vort}}
\newcommand{\FLSLA}{\Flength_{\mathrm{SLA}}}
\newcommand{\FLlsq}{\Flength_{\mathrm{lsq}}}

\begin{document}

\maketitle


\vspace{-10pt}
\noindent The following article appeared in \textit{Phys.~Fluids} 29, 115109 (2017) and may be found at \url{https://doi.org/10.1063/1.5012546}.
This article may be downloaded for personal use only. Any other use requires prior permission of the author and AIP Publishing.

\paragraph{Abstract} Direct numerical simulations of the incompressible Navier--Stokes
equations are {not feasible yet for most practical turbulent
  flows}. Therefore, dynamically less complex mathematical
formulations are necessary for coarse-grained simulations. In this
regard, eddy-viscosity models for Large-Eddy Simulation (LES) are
probably the most popular example thereof. This type of models
requires the calculation of a subgrid characteristic length which is
usually associated with the local grid size. For isotropic grids this
is equal to the mesh step. However, for anisotropic or unstructured
grids, {such as the pancake-like meshes that are often used to resolve
  near-wall turbulence or shear layers,} a consensus on defining the
subgrid characteristic length has not been reached yet despite the
fact that it can strongly affect the performance of LES models. In
this context, a new definition of the subgrid characteristic length is
presented in this work. {This flow-dependent length scale is based on
  the turbulent, or subgrid stress, tensor and its representations on
  different grids.}  The simplicity and mathematical properties
suggest that it can be a robust definition that minimizes the effects
of mesh anisotropies {on simulation results}. The performance of the
proposed subgrid characteristic length is successfully tested for
decaying isotropic turbulence and a turbulent channel flow using
artificially refined grids. Finally, a simple extension of the method
for unstructured meshes is proposed and tested for a turbulent flow
around a square cylinder. Comparisons with {existing subgrid
  characteristic length scales} show that the proposed definition is
much more robust with respect to mesh anisotropies and has a great
potential to be used in complex geometries where highly skewed
(unstructured) meshes are present.

\section{Introduction}

\label{intr}

The Navier--Stokes (NS) equations are an excellent mathematical model
for turbulent flows. {However, direct numerical simulations are
  not feasible yet for most practical turbulent flows, because the
  nonlinear convective term produces far too many scales of motion.}
Hence, in the foreseeable future, numerical simulations of turbulent
flows will have to resort to models of the small scales. The most
popular example thereof is Large-Eddy Simulation (LES). Briefly, the
LES equations arise from applying a spatial filter, with filter length
$\Flength$, to the NS equations, resulting in
\begin{equation}
\label{LESeq}
\partial_t \F{\vel} + ( \F{\vel} \cdot \grad ) \F{\vel} = \nu \lapl \F{\vel} - \nabla \F{p} - \nabla \cdot \tau \hspace{1mm}; \hspace{5mm} \nabla \cdot \F{\vel} = 0 .
\end{equation}
\noindent {Here, $\F{\vel}$ is the filtered velocity and $\tau =
  \F{\vel \otimes \vel} - \F{\vel} \otimes \F{\vel}$ is the subgrid
  stress (SGS) tensor that represents the effect of the unresolved
  scales.}  It is assumed that the filter $\vel \rightarrow \F{\vel}$
commutes with differentiation. {Since the SGS tensor, $\tau$,
  depends not only on the filtered velocity, $\F{\vel}$, but also on
  the full velocity field, $\vel$, we encounter a closure problem.  We
  thus have to approximate $\tau$ by a tensor depending only on the
  filtered velocity, \ie~$\tau \approx \tau ( \F{\vel} )$.}

Because of its inherent simplicity and robustness, the eddy-viscosity
assumption is by far the most used closure model,
\begin{equation}
\label{eddyvis}
\tau ( \F{\vel} ) \approx - 2 \nut \S ( \F{\vel} ) ,
\end{equation}
\noindent where $\nut$ denotes the eddy-viscosity and $\S ( \F{\vel} )
= 1/2 ( \grad \F{\vel} + \grad \F{\vel}\traspose )$ is the
rate-of-strain tensor.  Notice that $\tau ( \F{\vel} )$ is considered
traceless without loss of generality because the trace can be included
as part of the filtered pressure, $\F{p}$. Then, most of the existing
eddy-viscosity models can be expressed as follows: 
\begin{equation}
\label{eddyvis_template}
\nut = ( C_m \Flength )^2 D_m ( \F{\vel} ) ,
\end{equation}
\noindent where $C_m$ is the model constant, $\Flength$ is the subgrid
characteristic length and $D_m ( \F{\vel} )$ is the differential
operator with units of frequency associated with the model. {Here,
  no summation over $m$ is implied.}

In the last few decades research has primarily focused on either the
calculation of the model constant, $C_m$, or the development of more
appropriate model operators, $D_m ( \F{\vel} )$. {An example of
  the former is the approach proposed by Lilly~\cite{LIL67} to
  determine the model constant of the Smagorinsky model~\cite{SMA63}.}
{For an isotropic mesh, \ie~$\Flength=\Dx=\Dy=\Dz$, and under the
  assumption that the cutoff wave number $\kc = \pi / \Flength$ lies
  within the inertial range of a universal Kolmogorov spectrum, $E(k)
  = C_K \varepsilon^{2/3} k^{-5/3}$, a model's constant, $C_m$, can be
  found by assuming that its dissipation is equal to the turbulent
  kinetic energy dissipation, $\varepsilon$. In this way,
  Lilly~\cite{LIL67} obtained the Smagorinsky constant $C_S = ( 2/3
  C_K )^{3/4} / \pi$ (taking a value of $C_K \approx 1.58$ for the
  Kolmogorov constant~\cite{DON10} leads to $C_S \approx 0.17$).}

The classical Smagorinsky model~\cite{SMA63} has the disadvantage that
its differential operator, $D_m ( \F{\vel} ) = | \S ( \F{\vel} ) |$,
does not vanish near solid walls. Early attempts to overcome this
inherent problem of the Smagorinsky model made use of wall
functions~\cite{MOI82,PIO89}. {Later, Germano~\etal~\cite{GER91}
  proposed the dynamic procedure, in which the constant $C_m$ is
  computed with the help of the Jacobi identity (in least-squares
  sense), as was originally proposed by Lilly~\cite{LIL92}.} However,
this approach leads to highly variable coefficient fields with a
significant fraction of negative values for $\nut$. This can cause
numerical instability in simulations. Thus, averaging with respect to the
homogeneous direction(s) and {\itshape ad hoc} clipping {of} $\nut$
are, in general, necessary. Therefore, the original dynamic procedure
cannot be applied to geometrically complex flows without homogeneous
directions. Several attempts to overcome these intrinsic limitations
can be found in the literature: namely, the dynamic localization model
and the Lagrangian dynamic model were, respectively, proposed by
Ghosal~\etal~\cite{GHO95} and Meneveau~\etal~\cite{MEN96}. In the same
vein, Park~\etal~\cite{PAR06} introduced two global dynamic
approaches: a dynamic global model based on the Germano
identity~\cite{GER92} and a dynamic global model with two test filters
based on the “global equilibrium” between the viscous dissipation and
the SGS dissipation. Later, You and Moin~\cite{YOU07} presented a
dynamic global approach using only one test
filter. Tejada-Mart\'{i}nez and Jansen~\cite{TEJ04,TEJ06} proposed an
approach where the filter width ratio, the sole model parameter in a
dynamic Smagorinsky model, is computed dynamically too. To do so, they
assume scale invariance and make use of a secondary test filter.

{To construct models that vanish near solid walls, one can
  alternatively change the differential operator, $D_m ( \F{\vel} )$.
  Examples thereof are the WALE model~\cite{NIC99}, Vreman's
  model~\cite{VRE04b}, the QR model~\cite{VER11} and the
  $\sigma$-model~\cite{NIC11}.} This list can be completed with a
novel eddy-viscosity model proposed by Ryu and Iaccarino~\cite{RYU14}
and two eddy-viscosity models recently proposed by the authors of this
paper: namely, the S3PQR models~\cite{TRI14-Rbased} and the
vortex-stretching-based eddy-viscosity model~\cite{SIL17}.

Surprisingly, {in the LES community} little attention has been
paid to the computation of the subgrid characteristic length,
$\Flength$, which is also a key element of any eddy-viscosity model
[see Eq.~(\ref{eddyvis_template})]. Due to its simplicity and
applicability to unstructured meshes, nowadays the most widely used
approach to compute the subgrid characteristic length is the one
proposed by Deardorff~\cite{DEA70}, \ie~the cube root of the cell
volume. For a Cartesian grid it reads
\begin{equation}
\label{DeltaDeardorff}
\FLvol = ( \Dx \Dy \Dz )^{1/3} .
\end{equation}
Extensions of this approach for anisotropic grids where proposed by
Schumann~\cite{SCH75}, Lilly~\cite{LIL88} and
Scotti~\etal~\cite{SCO93}. It was found that for small anisotropies
{Deardorff's length scale is reasonably accurate, whereas
  corrections are required for highly anisotropic meshes, such as the
  pancake-like meshes that are often used to resolve near-wall
  turbulence or shear layers.} For instance, the following correction
was proposed by Scotti~\etal~\cite{SCO93},
\begin{equation}
\label{DeltaScotti}
\FLSco = f ( a_1 , a_2 ) \FLvol ,
\end{equation}
\noindent where $f ( a_1 , a_2 ) = \cosh \sqrt{ 4/27 [ ( \ln a_1 )^2 -
    \ln a_1 \ln a_2 + (\ln a_2)^2 ]}$ and $a_1 = \Dx/\Dz$, $a_2 =
\Dy/\Dz$, assuming that $\Dx \le \Dz$ and $\Dy \le \Dz$.
Nevertheless, it is {\itshape ``still assume(d), however, that the small
  scale limit of the simulation is in the Kolmogorov inertial
  sub-range in all directions. This is a serious limitation, as the
  most common reason for applying anisotropic resolution is an
  expectation of anisotropic and/or inhomogeneous turbulence,
  typically in regions close to a boundary''}~\cite{LIL88}. The
definition of $\Flength$ given in Eq.~(\ref{DeltaScotti}) was tested by
Scotti~\etal~\cite{SCO97} for forced isotropic turbulence using highly
anisotropic grids. They compared the correction factor $f ( a_1, a_2
)$ with the correction $f_{dyn} ( a_1, a_2)$ obtained by applying the
above-mentioned dynamic approach~\cite{GER91} to $C_m \FLvol f_{dyn} (
a_1, a_2 )$ (the dynamic approach is usually applied to find the model
constant, $C_m$). They reached the conclusion that the dynamic model
reproduces the correct trend for {pancake-like grids ($a_2=1, \Dx
  \ll \Dy = \Dz$), but fails for pencil-like ones ($a_1=a_2, \Dx = \Dy
  \ll \Dz$).} Another limitation of the approach proposed by
Scotti~\etal~\cite{SCO93} [see Eq.~(\ref{DeltaScotti})] is that it is
applicable only to structured Cartesian grids. To circumvent this,
Colosqui and Oberai~\cite{COL08} proposed an extension applicable to
unstructured meshes. They assume that the second-order structure
function satisfies Kolmogorov's hypotheses.

Alternative definitions of {the subgrid characteristic length scale,} $\Flength${,} include the maximum of the cell
sizes,
\begin{equation}
\label{Deltamax}
\FLmax = \max ( \Dx , \Dy , \Dz ) ,
\end{equation}
\noindent the $L^2$-norm of the tensor $\DelTen \equiv \diag ( \Dx,
\Dy, \Dz )$ divided by $\sqrt{3}$,
\begin{equation}
\label{DeltaL2}
\FLLtwo = \sqrt{(\Dx^2 + \Dy^2 + \Dz^2)/3} ,
\end{equation}
\noindent and the square root of the harmonic mean of the squares of
the grid sizes
\begin{equation}
\label{DeltaLapl}
\FLLapl = \sqrt{3/(1/\Dx^2+1/\Dy^2+1/\Dz^2)} ,
\end{equation}
\noindent which is directly related to the largest eigenvalue of the
discrete approximation of {the (negative) Laplacian,}
$-\lapl$. The first definition, $\FLmax$, was originally proposed in
the first presentation of the Detached-Eddy Simulation (DES) method by
Spalart~\etal~\cite{SPA97} as a safer and robust definition of
$\Flength$.

More recent definitions of {the subgrid characteristic length
  scale} are also found in the context of DES. Namely,
Chauvet~\etal~\cite{CHA07} introduced the concept of sensitizing
$\Flength$ {to the local velocity field.  In particular, they made
  $\Flength$ dependent on the orientation of the vorticity vector,
  ${\vort} = ( \vortx, \vorty, \vortz ) = \nabla \times {\vel}$,}
\begin{equation}
\label{Deltavort}
\FLvort = \sqrt{ ( \vortx^2 \Dy \Dz + \vorty^2 \Dx \Dz + \vortz^2 \Dx \Dy ) / | \vort |^2} ,
\end{equation}
\noindent with some minor corrections to prevent indeterminate forms
of type $0/0$ (see the original paper~\cite{CHA07} for details). The
formulation was subsequently generalized for unstructured meshes by
Deck~\cite{DEC12}. The definition of $\FLvort$ detects the alignment
of the vorticity vector, $\vort$, with an axis, \eg~if $\vort = ( 0 ,
0 , \vortz )$ then $\FLvort$ reduces to $\sqrt{\Dx \Dy}$. This
approach was motivated by the fact that $\FLmax$ results in an
excessive generation of SGS {dissipation} in the initial region of
shear layers typically resolved on highly anisotropic grids. In a DES
simulation, this results in an artificial delay of \KH instabilities
because the model switches from Reynolds-averaged Navier--Stokes (RANS) to LES mode further
downstream. However, like Deardorff's definition of $\Flength$ given
in Eq.~(\ref{DeltaDeardorff}), the definition given in
Eq.~(\ref{Deltavort}) may still involve the smallest of the grid
spacings. This may lead to very low values of eddy-viscosity.

To circumvent this problem, recently Mockett~\etal~\cite{MOC15}
proposed the following {flow-dependent subgrid characteristic
  length scale,}
\begin{equation}
\label{DeltaShu}
\FLMoc = \frac{1}{\sqrt{3}} \max_{n,m=1,\dots,8} | \isvector{l}_n - \isvector{l}_m | ,
\end{equation}
\noindent where $\isvector{l} = \vort/|\vort| \times \isvector{r}_n$
and $\isvector{r}_n$ ($n=1,\dots,8$ for a hexahedral cell) are the
locations of the cell vertices. The quantity $\FLMoc$ represents the
diameter of the set of cross-products points, $\isvector{l}_n$,
divided by $\sqrt{3}$. In the above-described case with $\vort = ( 0,
0, \vortz )$ it reduces to $\FLMoc = \sqrt{ ( \Dx^2 + \Dy^2 ) / 3 }$.
Therefore, it is ${\mathcal O} ( \max \{ \Dx, \Dy \} )$ instead of $\FLmax
= \Dz$ (for the typical situation where $\Dz > \Dx$ and $\Dz > \Dy$)
or $\FLvort = \sqrt{ \Dx \Dy }$. Thus, \textit{``unlike $\FLvort$ the
  definition}~(\ref{DeltaShu}) \textit{never leads to a strong effect of the
  smallest grid-spacing on the subgrid-scale $\Flength$ even though it
  achieves the desired decrease compared to the $\FLmax$ definition in
  the quasi-2D flow regions treated on strongly anisotropic
  grids.''}~\cite{SHU15}.

More recently, Shur~\etal~\cite{SHU15} proposed to modify the
definition of $\FLMoc$ given in Eq.~(\ref{DeltaShu}) by introducing a
nondimensional function $0 \le F_{KH} ( VTM ) \le 1$, resulting in the
Shear Layer Adapted (SLA) subgrid scale
\begin{equation}
\label{DeltaSLA}
\FLSLA = \FLMoc F_{KH} ( VTM ) ,
\end{equation}
\noindent where the Vortex Tilting Measure (VTM) is given by
\begin{equation}
VTM = \frac{| ( \S \cdot \vort ) \times \vort |}{\vort^2 \sqrt{-\Q_\traceless{\S}}} ,
\end{equation}
\noindent where $\traceless{\S}$ is the traceless part of the
rate-of-strain tensor, $\S = 1/2 ( \nabla \F{\vel} + \nabla
\F{\vel}\traspose )$, \ie~$\traceless{\S} = \S - 1/3 \tr{\S}
\Identity$. Note that for incompressible flows $\tr{\S} = \diver
\F{\vel} = 0$, therefore, $\traceless{\S} = \S$. Finally $\QA$ denotes
the second invariant of a second-order tensor $\A$, $\QA = 1/2 \{
\trpow{2}{\A} - \tr{\A^2} \}$. {The vortex tilting measure is
  bounded, $0 \le VTM \le 1$, and} it takes zero value when the
vorticity is aligned with an eigenvector of $\S$ with eigenvalue
$\lambda_i$, \ie~$\S \vort = \lambda_i \vort$. Therefore, the VTM can
be viewed as a measure of how much the rate-of-strain tensor tilts the
vorticity vector towards another direction. Finally, the function
$F_{KH}$ is aimed at unlocking the \KH instability in the initial part
of shear layers. Different functions $F_{KH}$ were proposed by
Shur~\etal~\cite{SHU15} with the basic requirements that $0 \leq
F_{KH} ( VTM ) \leq 1$ and $F_{KH} ( 0 ) = 0$ and $F_{KH} ( 1 ) = 1$.

{Despite the above-mentioned length scales, so far no consensus
  has been reached on how to define the subgrid characteristic length
  scale, particularly when considering anisotropic or unstructured
  grids. In this work, we therefore propose a new flow-dependent
  subgrid characteristic length scale that is based on the subgrid
  stress tensor, $\tau$, and its representations on different grids.
  This simple and robust definition of $\Flength$ reduces the effect
  of mesh anisotropies on the performance of SGS models.}

{The structure of this paper is as follows. In
  Section~\ref{properties} all the above-mentioned definitions of the
  subgrid characteristic length are compared and classified on the
  basis of a list of desirable properties. These properties are based
  on physical, numerical and/or practical arguments.  Then, within
  this framework, a new subgrid characteristic length, based on the
  Taylor-series expansion of the SGS tensor in the computational
  space, is proposed in Section~\ref{building}. Moreover, in
  Section~\ref{unstr} a simple extension of this length scale for
  unstructured grids is proposed. In Section~\ref{results} the newly
  proposed length scale is tested in wall-resolved large-eddy
  simulations on highly anisotropic structured grids (test cases:
  decaying isotropic turbulence and a plane-channel flow) and
  unstructured grids (test case: turbulent flow around a square
  cylinder), confirming that it is a robust definition that reduces
  the effects of mesh anisotropies on the performance of LES models.
  Finally, relevant results are summarized and conclusions are given
  in Section~\ref{conclusions}.}


\begin{table}
\small
\centering
\begin{tabular}{lccccccccc}
\toprule
  & $\FLvol$          & $\FLSco$        & $\FLmax$      & $\FLLtwo$     & $\FLLapl$     & $\FLvort$    & $\FLMoc$  & $\FLSLA$  & $\FLlsq$     \\
Formula      & Eq.~(\ref{DeltaDeardorff}) & Eq.~(\ref{DeltaScotti}) & Eq.~(\ref{Deltamax}) & Eq.~(\ref{DeltaL2}) & Eq.~(\ref{DeltaLapl}) & Eq.~(\ref{Deltavort}) & Eq.~(\ref{DeltaShu})        & Eq.~(\ref{DeltaSLA})           & Eq.~(\ref{DeltaLsq}) \\
\midrule
{\bfseries P1}     & Yes                      & Yes                    & Yes                 & Yes                & Yes                & Yes                  & Yes     & Yes                   & Yes \\
{\bfseries P2}     & Yes                      & Yes                    & Yes                 & Yes                & Yes                & Yes                  & Yes     & Yes                   & Yes \\
{\bfseries P3}     & No                       & No                     & No                  & No                 & No                 & Yes                  & Yes     & Yes                   & Yes \\
{\bfseries P4}     & Yes                      & No                     & No\footA                 & No                 & No\footA                & No\footB                 & Yes     & Yes                   & Yes \\
{\bfseries P5}     & Low                      & Medium                 & Low                 & Low                & Low                & Medium               & High    & High                  & Low \\
\bottomrule
\end{tabular}
\raggedright \footA Possible with some adaptations. \\
\raggedright \footB Deck~\cite{DEC12} proposed a generalization for unstructured
meshes.
\caption{Properties of different definitions of the subgrid
  characteristic length, $\Flength$. Namely, {\bfseries P1}: positive
  ($\Flength \ge 0$), local and frame invariant; {\bfseries P2}: bounded,
  \ie~given a structured Cartesian mesh where $\Dx \le \Dy \le \Dz$,
  $\Dx \le \Flength \le \Dz$; {\bfseries P3}: sensitive to {the local
    flow field}; {\bfseries P4}: applicable to unstructured meshes; {\bfseries
    P5}: computational cost.}
\label{properties_Delta}
\end{table}

\newcommand{\wall}[1]{{\mathcal O}( y^{#1} )}

\section{Properties of the subgrid characteristic length}

\label{properties}

Starting from the classical {Smagorinsky model~\cite{SMA63},} many
eddy-viscosity models [see Eq.~(\ref{eddyvis})] have been proposed
(see the work of Trias~\etal~\cite{TRI14-Rbased}, for a recent review). The
definition of $\nut$ given in Eq.~(\ref{eddyvis_template}) provides a
general template for most of them. Therefore, a subgrid characteristic
length, $\Flength$, which is commonly associated with the local grid
size, is required. Hence, for isotropic grids $\Flength$ is equal to
the mesh size, \ie~$\Flength=\Dx=\Dy=\Dz$. However, for anisotropic or
unstructured grids, a consensus on defining the subgrid characteristic
length has not been reached yet. Despite the fact that in some
situations it may provide very inaccurate results, three and a half
decades later, the approach proposed by Deardorff~\cite{DEA70},
\ie~the cube root of the cell volume {[see
  Eq.~(\ref{DeltaDeardorff})]}, is by far the most widely used.

Alternative methods to compute {the subgrid characteristic length
  scale, }$\Flength${,} have been reviewed in Section
\ref{intr}. They are classified in Table~\ref{properties_Delta}
according to a list of desirable properties for a (proper) definition
of $\Flength$. These properties are based on physical, numerical,
and/or practical arguments. Namely, the first property, denoted as
{\bfseries P1}, entails both positiveness and locality. Although from a
physical point of view negative values of $\nut$ may be justified with
the backscatter phenomenon, from a numerical point of view, the
condition $\nut \ge 0$ is, in general, considered appropriate because
it guarantees stability. Further, the LES equations should be Galilean
invariant. In order to preserve this physical principle, the
flow-dependent definitions of $\Flength$ (see property {\bfseries P3} below)
are based on invariants derived from the gradient of the resolved
velocity field, $\G \equiv \nabla \F{\vel}$. In doing so, the condition
of locality is also achieved. From a practical point of view, locality
is a desirable feature especially if the model is aimed to be applied
in complex flows. The second property ({\bfseries P2}) requires that
$\Flength$ is properly bounded, \ie~given a structured Cartesian mesh
where $\Dx \le \Dy \le \Dz$, we need $\Dx \le \Flength \le \Dz$. These
first two properties {\bfseries P1} and {\bfseries P2} are achieved by all the
length scales shown in Table~\ref{properties_Delta}. The third
property ({\bfseries P3}) classifies the methods to compute $\Flength$ in
two families: those that solely depend on geometrical properties of
the mesh, and those that are also dependent on the local flow
topology, $\ie$~the velocity gradient, $\G$.

{Assuming that the grid is Cartesian, we can express the subgrid
  characteristic length scales that are fully mesh-based in terms of
  the properties of the following second-order diagonal tensor,}
\begin{equation}
\label{DelTen_def}
{\DelTen \equiv \diag ( \Dx , \Dy , \Dz ).}
\end{equation}
{We take $\Dx \le \Dy \le \Dz$ without
  loss of generality. Indeed the aforementioned mesh-based length
  definitions can be written as}
\begin{equation}
\label{DeltaDeardorff2}
{\FLvol = \R_\DelTen^{1/3} , \hphantom{aa} 
\FLSco = f ( a_1 , a_2 ) \R_\DelTen^{1/3} , \hphantom{aa} 
\FLmax = \lambda_1^{\DelTen} , \hphantom{aa} 
\FLLtwo = \sqrt{ \frac{\trace ( \DelTen^2 )}{3} } = \sqrt{ \frac{\P_\DelTen^2 - 2 \Q_\DelTen}{3}},}
\end{equation}
{\noindent where the correction function $f ( a_1 , a_2 )$ was
  defined in Eq.~(\ref{DeltaScotti}) and $a_1 = \Dx/\Dz$, $a_2
  =\Dy/\Dz$. Moreover, $P_{\DelTen} = \tr{\DelTen}$, $Q_{\DelTen} =
  1/2 \{ \trpow{2}{\DelTen} - \tr{\DelTen^2} \}$ and $R_{\DelTen} =
  \Det(\DelTen) = 1/6 \{ \trpow{3}{\DelTen} - 3
  \tr{\DelTen}\tr{\DelTen^2} + 2 \tr{\DelTen^3} \} $ represent the
  first, second and third invariant of the second-order tensor
  $\DelTen$, respectively. The three eigenvalues, $\lambda_1^\DelTen
  \ge \lambda_2^\DelTen \ge \lambda_3^\DelTen$, of $\DelTen$ are
  solutions of the characteristic equation, $\Det ( \lambda \Identity
  - \DelTen ) = \lambda^3 - P_{\DelTen} \lambda^2 + Q_{\DelTen}
  \lambda - R_{\DelTen} = 0$.}

{The aforementioned characteristic length scale definitions that depend on
both the mesh and the flow topology can also be expressed in terms of
invariants, namely the invariants of $\DelTen$, the velocity gradient
$\G$ and the invariants of $\G$'s symmetric and anti-symmetric
part. {To this end, note} that the vorticity vector ${\vort} = ( \vortx, \vorty,
\vortz ) = \nabla \times \F{\vel}$ can be expressed in terms of the
rate-of-rotation tensor, $\O = 1/2 ( \G - \G\traspose )$, as $\omega_k
= - \epsilon_{ijk} \O_{ij}$ where $\epsilon_{ijk}$ is the Levi--Civita
symbol. {Here, summation over repeated indices is implied.} Finally, the symmetric part of $\G$ is the rate-of-strain
tensor, $\S = 1/2 ( \G + \G\traspose )$. These are the flow-dependent
quantities required to compute the definitions of $\Flength$ of
Chauvet~\etal~\cite{CHA07} [see the definition of $\FLvort$ given in
Eq.~(\ref{Deltavort})], Mockett~\etal~\cite{MOC15} [see the definition
of $\FLMoc$ given in Eq.~(\ref{DeltaShu})] and the modification $\FLSLA$
proposed by Shur~\etal~\cite{SHU15} [see Eq.~(\ref{DeltaSLA})].}

The last two {desirable} properties {of subgrid characteristic length scales} are of practical interest. Namely, property
{\bfseries P4} refers to the applicability of the method for unstructured
meshes. Among the definitions of $\Flength$ that do not depend on the
local flow, only the approach of Deardorff~\cite{DEA70} can be
straightforwardly used for unstructured grids. Recent flow-dependent
definitions of $\Flength$ are potentially applicable for unstructured
grids, although some of them have a relatively high computational
cost. In this regard, to complete the list of properties, it is also
desirable that the definition of $\Flength$ is well conditioned and
has a low (or moderate) computational cost (property {\bfseries P5}). In
this respect, flow-dependent definitions of $\Flength$ may be
problematic, having a significantly higher computational
cost. Moreover, they require special attention for indeterminate forms
of type $0/0$. 

The new definition of the subgrid characteristic
length, $\FLlsq$, which is presented in the next section matches all
the above-mentioned properties with an inherent simplicity and a
moderate computational cost.


\section{Building a new subgrid characteristic length}

\label{building}

Several approaches to compute the subgrid characteristic length,
$\Flength$, can be found in the literature (see
Section~\ref{intr}). Their properties have been analyzed and compared
in Section~\ref{properties} (see
Table~\ref{properties_Delta}). {As remarked before, despite these
  existing length scales, no consensus has been reached on how to
  define the subgrid characteristic length scale, particularly for
  (highly) anisotropic or unstructured grids. In this section, we
  therefore propose a new flow-dependent subgrid characteristic length
  scale that is based on the subgrid stress tensor, $\tau$, and its
  representations on different grids.}

{The subgrid characteristic length, $\Flength$, appears in a
  natural way when we consider the lowest-order approximation of the
  subgrid stress tensor, $\tau = \F{\vel \otimes \vel} - \F{\vel}
  \otimes \F{\vel}$, \ie~the unclosed term in the filtered
  Navier--Stokes equations, Eq.~(\ref{LESeq}). The approximation of the
  subgrid stress is obtained by approximating the residual velocity $u'
  = u-\F{u}$.}
  
{To start, we restrict ourselves to one spatial direction
  and consider a box filter. The residue of the box filter can be
  related to the error of the midpoint rule for numerical integration,
  denoted by $\epsilon$ here. We have $\F{u} = \int_{x-\Dx/2}^{x+\Dx/2}
  u(x) \,dx = \Dx u(x) + \epsilon$ with $\epsilon = \frac{\Dx^3}{24}
  \partial_{x} \partial_x u(c)$ where $c$ lies somewhere in between
  $x-\Dx/2$ and $x+\Dx/2$. An expression for the residue of the
  one-dimensional box filter is then obtained by dividing this error
  by $-\Delta x$ and adding $u$. Thus to lowest order we get $u'(x) = -
  \frac{\Dx^2}{24} \partial_x \partial_x u (x) + {\mathcal{O}}(\Delta
  x^4)$.}

{On a three-dimensional, isotropic grid, \ie~$\Delta = \Delta x =
  \Delta y = \Delta z$, the above approximation of the residue becomes
  $u' = - \frac{\Flength^2}{24} \nabla \cdot \nabla u +
         {\mathcal{O}}(\Flength^4)$. With the help of this approximation
         it can be shown that the subgrid stress tensor is given
         by~\cite{CLA79}}
\begin{equation}
\label{GradMod}
{\tau ( \F{\vel} ) = \frac{\Flength^2}{12} \GGt + {\mathcal O} ( \Flength^4 ) .}
\end{equation}
{The leading-order term of Eq.~(\ref{GradMod}) is the gradient
  model proposed by Clark~\etal~\cite{CLA79}, where $\Flength$ denotes
  the filter length. Equation (\ref{GradMod}) has been derived for the
  box filter. However, it can be shown that the same result is
  obtained for any convolution filter having a symmetric
  kernel~\cite{WIN01}.}

{We stress that in the above derivation the grid is assumed to be
isotropic, that is $\Delta = \Delta x = \Delta y = \Delta z$. For an
anisotropic grid, we can postulate that the lowest-order approximation
of the subgrid stress also provides us with $\tau ( \F{\vel} ) \approx
\frac{\Flength^2}{12} \GGt$, that is, the approximation (1) depends
quadratically on the velocity gradient, (2) is given by a symmetric
tensor, (3) is invariant under a rotation of the coordinate system,
and (4) is proportional to $\Flength^2$. Here, however, we do not yet
know how to define the filter length, $\Flength$, because the grid is
anisotropic. For the gradient model, however, we can define the
filter length by mapping the anisotropic mesh onto an isotropic mesh.
Therefore we consider the coordinate transformation $\hat{x} =x/\Dx$,
$\hat{y} =y/\Dy$ and $\hat{z} =z/\Dz$. Expanding the subgrid stress
as before, but now in the new, isotropic, coordinate system
$\hat{x},\hat{y},\hat{z}$ and applying the chain rule for
differentiation we obtain}
\begin{equation}
\label{AniGradMod}
{\tau ( \F{\vel} ) = \frac{1}{12} \GD \GD\traspose + {\mathcal O} ( \Flength^4 ) .}
\end{equation}
{Here, the velocity gradient on the anisotropic grid is defined as}
\begin{equation}
{\GD \equiv \G \DelTen ,}
\end{equation}
{where $\DelTen$ is the second-order tensor containing the mesh information given by Eq.~(\ref{DelTen_def}).}
{Equation (\ref{AniGradMod}) does not require an explicit
  definition of the filter length, $\Flength$. In fact the filter
  length is hidden in $\GD$ and is not represented by a scalar but by
  the tensor $\DelTen$. Since both
  Eq.~(\ref{GradMod}) and Eq.~(\ref{AniGradMod}) represent the
  lowest-order approximation of the subgrid stress, we can equate them
  and thus define the filter length $\Flength$ in Eq.~(\ref{GradMod})
  for anisotropic meshes. Here it may be remarked that we equate
  tensors; hence the equality is to be understood in least-square
  sense. This leads to the following flow-dependent definition of
  $\Flength$,}
\begin{equation}
\label{DeltaLsq}
{\FLlsq = \sqrt{\frac{\GD \GD\traspose : \GGt}{\GGt : \GGt}} .}
\end{equation}
{We first remark that this length scale reduces to $\Flength$ on
  an isotropic mesh. Secondly, since $\FLlsq$ is formally based on the
  lowest-order approximation of the subgrid stress, we see it as a
  generic way to define the filter length. It can thus be applied in
  any turbulence model, not only in eddy-viscosity models,
  Eq.~(\ref{eddyvis_template}). With respect to the properties
  discussed in Section~\ref{properties}, the characteristic length
  scale given by Eq.~(\ref{DeltaLsq}) depends on the velocity
  gradient, $\G$.  Therefore, it is locally defined and frame
  invariant ({\bfseries P1}).  Moreover, $\FLlsq$ is obviously sensitive to
  flow orientation (property {\bfseries P3}).}

{Furthermore, it may be noted that the numerator in Eq.~(\ref{DeltaLsq})
can be viewed as the Frobenius norm of the tensor $\G\traspose\G
\DelTen$, \ie~$\GD \GD\traspose : \GGt = \trace(\GD\GD\traspose\GGt) =
\trace(\G\DelTen^2\G\traspose\GGt) = \trace(\DelTen\GtG
(\DelTen\GtG)\traspose) = \DelTen\GtG:\DelTen\GtG$. Moreover, $\GGt :
\GGt = \trace(\GGt\GGt) = \trace(\GtG\GtG) = \GtG : \GtG$, so we can
also express $\FLlsq$ as}
\begin{equation}
{\FLlsq = \sqrt{\frac{\DelTen\GtG : \DelTen\GtG}{\GtG : \GtG}} .}
\end{equation}

From this definition it is obvious that $\FLlsq$ is positive and well
bounded (properties {\bfseries P1} and {\bfseries P2}). Its applicability for
unstructured meshes (property {\bfseries P4}) relies on the proper
adaptation of the tensor $\DelTen$ (see Section~\ref{unstr}).
Regarding property {\bfseries P5}, the computational cost of $\FLlsq$ is
relatively small {when compared to the other flow-dependent
  (property {\bfseries P3}) length scales discussed in this paper} and
special attention is only required for indeterminate forms of type
$0/0$. 

{The inherent simplicity and mathematical properties of the
  proposed length scale, as well as its basis in representations of
  the subgrid stress tensor on different grids suggest that it can be
  a robust definition that minimizes the effects of mesh anisotropies
  on the performance of LES models.} {Note that the definition of
  $\FLlsq$ provided in Eq.~(\ref{DeltaLsq}) was already presented and
  partially evaluated during the Stanford CTR Summer Program
  2016~\cite{SILTRI16-CTR} and the CEAA'16
  conference~\cite{TRI16CEAA}.}

{To get a better understanding of $\FLlsq$, we consider several
  special cases. First of all, as remarked before, this length scale
  reduces to $\Flength$ on an isotropic mesh. Secondly,} for purely
rotating flows, \ie~$\S=\istensor{0}$ and $\G = \O$, $\FLlsq$ reduces
to
\begin{equation}
\label{Dlsq_rot}
\FLlsq = \sqrt{ \frac{\vortx^2 ( \Dy^2 + \Dz^2 ) + \vorty^2 ( \Dx^2 + \Dz^2 ) + \vortz^2 ( \Dx^2 + \Dy^2 )}{ 2 | \vort |^2}} ,
\end{equation}
\noindent which resembles the definition of $\FLvort$ proposed by
Chauvet~\etal~\cite{CHA07} given in Eq.~(\ref{Deltavort}). Actually,
similar to the definition of $\Flength$ proposed by
Mockett~\etal~\cite{MOC15} given in Eq.~(\ref{DeltaShu}), $\FLlsq$ is
${\mathcal O} ( \max \{ \Dx, \Dy \} )$ instead of $\FLvort = \sqrt{ \Dx
  \Dy }$. Therefore, it also avoids a strong effect of the smallest
grid-spacing. 
\begin{figure}[!t]
\centering{
  \includegraphics[height=0.6969\textwidth,angle=-90]{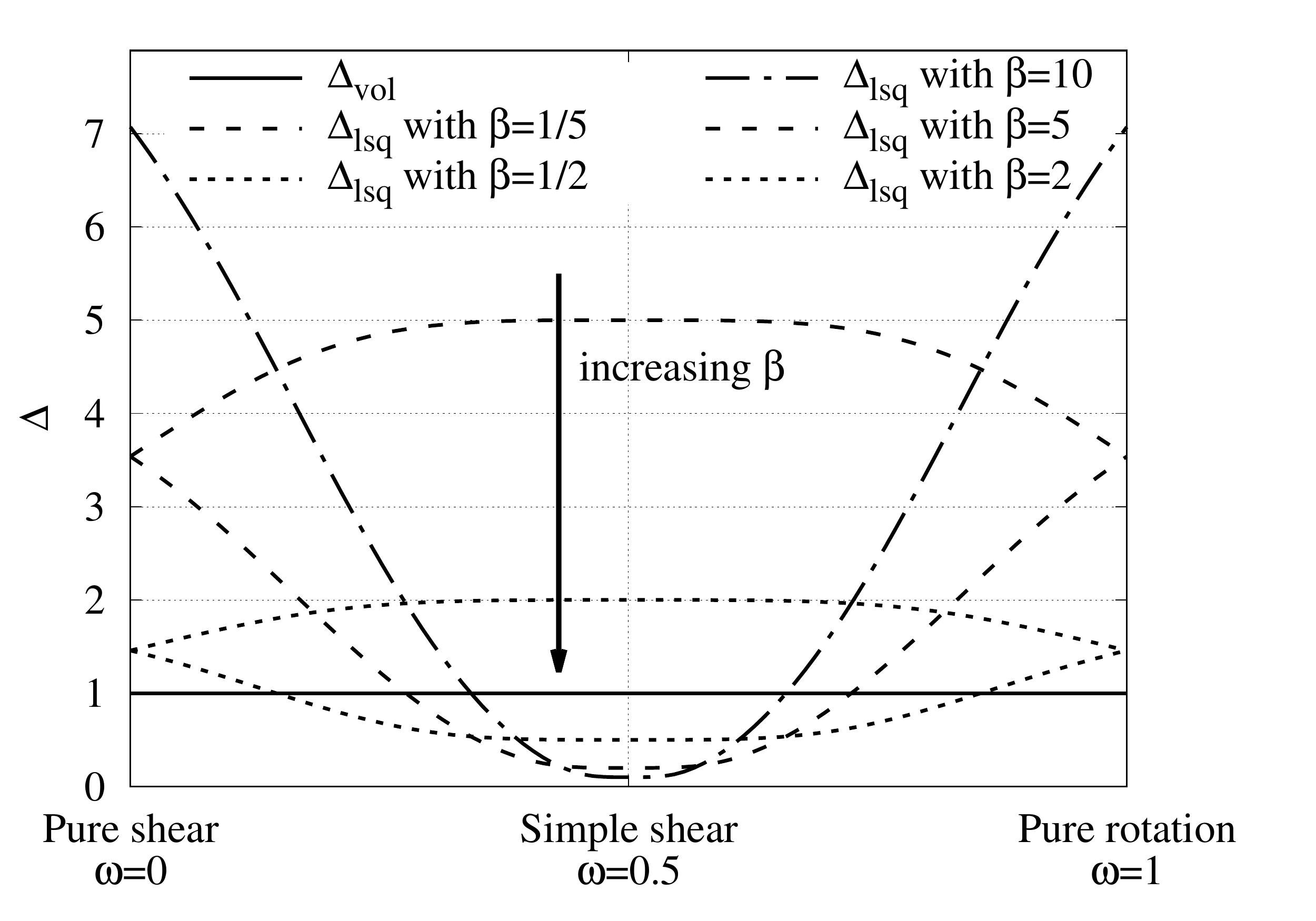}
}
\caption{Comparison between $\FLlsq$ and $\FLvol$ for the simple 2D
  flow defined in Eq.~(\ref{simpleflow}) with different values of
  $\beta=\{1/5,1/2,2,5,10\}$.}
\label{simple_flows}
\end{figure}
Finally, results obtained for a simple 2D mesh and flow,
\begin{equation}
\label{simpleflow}
\DelTen = \left( \begin{array}{cc} \beta & 0 \\ 0 & \beta^{-1} \end{array} \right) , \hspace{7mm}
\G = \left( \begin{array}{cc} 0 & 1 \\ 1-2\omega & 0 \end{array} \right) ,
\end{equation}
\noindent are displayed in Figure~\ref{simple_flows}. Notice that the
size of the control volume remains equal to unity; therefore, $\FLvol
= 1$, regardless of the value of $\beta$. On the other hand, values of
$\omega$ in Figure~\ref{simple_flows} range from a pure shear flow
($\omega=0$) to a simple shear flow ($\omega=1/2$), to a pure rotating
flow ($\omega=1$). For the two limiting situations $\FLlsq = \sqrt{(
  \beta^2 + \beta^{-2} )/2}$ whereas for $\omega=1/2$ it reads $\FLlsq
= \beta^{-1}$. Recalling that in the particular case $\Dx=\beta$ and
$\Dy=\beta^{-1}$, $\FLlsq = \sqrt{(\Dx^2 + \Dy^2)/2}$ for $\omega=0$
(pure shear) and $\omega=1$ (pure rotation), whereas $\FLlsq = \Dy$
for the simple shear flow with $\omega=1/2$. The latter corresponds
quite well with the typical quasi-2D grid-aligned flow in the initial
region of a shear layer. As it could be expected, the computed
$\FLlsq$ is equal to the grid size in the direction orthogonal to the
shear layer. The pure rotating flow ($\omega = 1$) is just a
particular case of Eq.~(\ref{Dlsq_rot}) with $\vortx=\vorty=0$ and
$\vortz=1$.

\begin{figure}[!t]
\centering{
  \includegraphics[height=0.6969\textwidth,angle=-90]{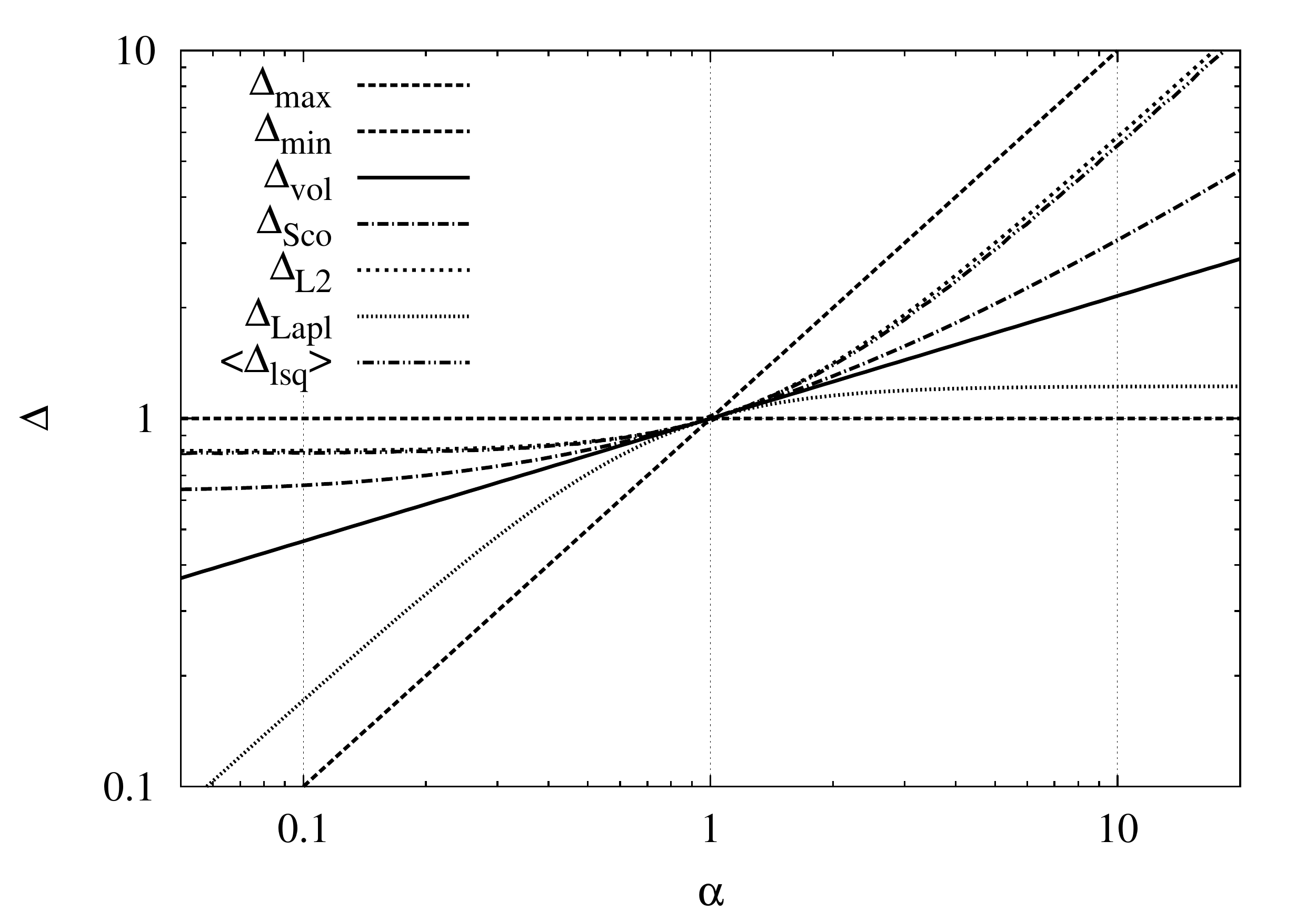}
}
\caption{Scaling of different definitions of $\Flength$ for a
  Cartesian mesh with $\Dx=\Dz=1$ and $\Dy=\alpha$. Average results of
  $\FLlsq$ have been obtained from a large sample of random traceless
  velocity gradient tensors.}
\label{scaling}
\end{figure}

In order to study in more detail the effect of mesh anisotropies for
different definitions of $\Flength$, let us consider a Cartesian mesh
with $\Dx=\Dy=1$ and $\Dz=\alpha$. In this case, the
geometry-dependent definitions of $\Flength$ result in
\begin{align}
\begin{split}
\FLvol &= \alpha^{1/3}, \hspace{5mm} \FLSco = f(\min(\alpha,\alpha^{-1}),\min(1,\alpha^{-1})), \\
\FLmax &= \max(1,\alpha), \hspace{5mm} \FLLtwo = \sqrt{\frac{2+\alpha^2}{3}}, \hspace{5mm} \FLLapl = \sqrt{\frac{3\alpha^2}{2\alpha^2+1}} .
\end{split}
\end{align}
These functions are displayed in Figure~\ref{scaling} using a log-log
scale. Values of $\alpha > 1$ correspond to pencil-like meshes {($\Dx = \Dy \ll \Dz$) }whereas
values of $\alpha < 1$ correspond to pancake-like meshes{ ($\Dx \ll \Dy = \Dz$)}. Averaged
results of $\FLlsq$ are also displayed; they have been obtained from a
large sample of random traceless velocity gradient tensors,
$\G$. Notice that this simple random procedure was able to produce
fairly good predictions to determine the model constant, $C_m$, for
different SGS models~\cite{TRI14-Rbased}. Among all the
geometry-dependent definitions, for the mesh considered here, the
closest to $\avgtime{\FLlsq}$ by far is the definition of $\FLLtwo$
given in Eq.~(\ref{DeltaL2}). Actually, for very simple flow
configurations such as pure shear or pure rotation, $\FLlsq$ reduces
to $\FLLtwo$. The second closest is the correction proposed by
Scotti~\etal~\cite{SCO93} [see Eq.~(\ref{DeltaScotti})]. Finally,
$\FLLapl$ is the only definition that predicts values of $\Flength$
smaller than the classical Deardorff definition, $\FLvol$.


\section{Jacobian-based extension for unstructured meshes}

\newcommand{\Gmat}{\boldsymbol{G}}
\newcommand{\Jmat}{\boldsymbol{J}}
\newcommand{\Gcomp}{\boldsymbol{\mathcal{G}}}
\newcommand{\Gcompcoeff}{\mathcal{G}}
\newcommand{\Gmatcoeff}{{G}}
\newcommand{\Jcoeff}{\mathcal{J}}

\label{unstr}

{In Section~\ref{building}, a new method to compute the subgrid
  characteristic length has been proposed. Although it has been
  derived in the context of Cartesian meshes, the idea can be extended
  to unstructured meshes by noticing that it basically consists in
  projecting the leading term of the Taylor series expansion of $\tau$
  [see Eq.~(\ref{AniGradMod})] onto the basic gradient model [see
  Eq.~(\ref{GradMod})]. 
}

{For non-uniform Cartesian grids we considered the coordinate
  transformation $\hat{x} =x/\Dx$, $\hat{y} =y/\Dy$ and $\hat{z}
  =z/\Dz$. This led to a new, isotropic, coordinate system
  $\hat{x},\hat{y},\hat{z}$. Then, applying the chain rule for
  differentiation yielded the approximation of the subgrid stress tensor of Eq.~(\ref{AniGradMod}). More generally,
  let $\xi_i(\x_i)$ be a monotonic differentiable function which
  defines a mapping from the physical space in the $i$-direction,
  $\x_i$, to the so-called computational space, $\xi_i$. Using the
  chain rule we obtain}
\begin{equation}
{\frac{\partial \phi}{\partial x_i} = \frac{\partial \phi}{\partial \xi_i} \frac{\ud \xi_i}{\ud x_i} = \frac{1}{J_i} \frac{\partial \phi}{\partial \xi_i} ,}
\end{equation}
\noindent {where $J_i$ is the Jacobian of the transformation $x_i
  \rightarrow \xi_i$. Here, no summation over $i$ is implied.}
Recalling that $[\G]_{ij} = \partial u_i / \partial x_j$, the leading
term of $\tau$ can be written more compactly as follows:
\begin{equation}
\label{GenGradMod}
\tau = \frac{1}{12} \G_\xi \G_\xi\traspose + {\mathcal O} ( \cFlength^4 )
,
\end{equation}
where the gradient in the mapped space $\isvector{\xi}$ is represented by
\begin{equation}
\G_\xi = \G \J
\end{equation}
and $\J$ is the Jacobian of the transformation
$\isvector{x} \rightarrow \isvector{\xi}$. Notice that this first term
is generic for all practical filters~\cite{WIN01} in the context of
LES, \ie~filters with a Fourier transform starting with
$\foutrans{G}(k) = 1 - k^2 \cFlength /2 + {\mathcal O} (k^4)$. At the
discrete level, for a Cartesian grid the filter length in each
direction is taken equal to the mesh size in the same direction,
\ie~$\cFlength_i = \Dx_i$. In this case, $\J = \DelTen$ and $\G_\xi =
\GD = \G \DelTen$; therefore, the general expression given in
Eq.~(\ref{GenGradMod}) reduces to Eq.~(\ref{AniGradMod}) for
non-uniform Cartesian meshes and to the well-known gradient
model~\cite{CLA79} given in Eq.~(\ref{GradMod}) for uniform grid
spacings.

At this point, it becomes clear that the extension of the new subgrid
characteristic length $\FLlsq$ [see Eq.~(\ref{DeltaLsq}) in
Section~\ref{building}] for unstructured meshes relies on the
computation of the Jacobian, $\J$, on such grids. It is important to
note that the gradient tensor, $\G$, is actually being computed in any
LES code. Below, the method to compute the Jacobian, $\J$, is solely
based on the discrete gradient operator; therefore, it can be easily
applied to any existing code. Namely, using matrix-vector notation,
the discrete gradient operator is given by a block matrix
\begin{equation}
\Gmat \scafield_h = \left( \begin{array}{c} \Gmat_x \\ \Gmat_y \\ \Gmat_z \end{array} \right) \scafield_h ,
\end{equation}
\noindent where $\scafield_h = ( \sca_1, \sca_2, ...,
\sca_\pvlength)\traspose \in \real^{\pvlength}$, $\pvlength$ is the
number of unknowns in our domain and $\Gmat_x$, $\Gmat_y$ and
$\Gmat_z$ represent the discrete gradient operator for each spatial
direction. 

As a preview of things, we first consider the
discretization of the gradient operator, $\Gmat$, in one spatial
direction with periodic boundary conditions. Let us consider three
values of a smooth function $\sca (x)$: $\sca_{i-1} = \sca(x_{i-1})$,
$\sca_{i} = \sca(x_{i})$ and $\sca_{i+1} = \sca(x_{i+1})$ with
$x_{i-1} = x_i - \Dx$ and $x_{i+1} = x_i + \Dx$. By a simple
combination of Taylor series expansions of $\sca(x)$ around $x=x_i$,
the following well-known second-order accurate approximation of the
derivative follows
\begin{equation}
\frac{\partial \sca(x_i)}{\partial x} \approx \frac{\sca_{i+1} - \sca_{i-1}}{2 \Dx} .
\end{equation}
\noindent Then with a uniformly meshed periodic direction, $\Gmat_x$
results into a skew-symmetric circulant matrix of the form
\begin{equation}
\Gmat_x = \frac{1}{2 \Dx} \Circ ( 0, 1, 0, \cdots, 0, -1 ) .
\end{equation}
\noindent Thus, eigenvalues of $\Gmat_x$ lie on the imaginary axis,
$\lambda_k^{\Gmat_x} \in \imag$. Then, the eigenvalues can be easily
bounded with the help of the Gershgorin circle theorem,
\ie~$|\lambda_{k}^{\Gmat_x}| \le 1/\Dx$. Notice that the upper bound
exactly corresponds to the Jacobian, $J_x = 1 / \Dx$, of the mapping
from the physical to the computational space {for Cartesian
  grids}. {This idea can be extended to any grid or numerical
  method if we consider that, at the discrete level, the Jacobian,
  $\Jmat$, is as a diagonal matrix}
\begin{equation}
\Jmat \equiv \left( \begin{array}{ccc} \Jmat_x & & \\ & \Jmat_y & \\ & & \Jmat_z \end{array} \right) ,
\end{equation}
\noindent which, {similar to the Cartesian case,} guarantees that the
spectral norm of the gradient in the so-called computational space,
$\Gcomp \equiv \Jmat \Gmat = ( \Gcomp_x, \Gcomp_y, \Gcomp_z
)\traspose$ is equal to or smaller than unity, \ie~$\| \Gcomp \|_2 \le
1$. {This condition can easily be realized} by using the
Gershgorin circle theorem. Namely,
\begin{equation}
|\lambda_{i}^{\Gcomp_x} - \Gcompcoeff^{x}_{ii}| \le \sum_{j \neq i} | \Gcompcoeff^x_{ij} | \hspace{5mm} \text{where} \hspace{5mm} \Gcompcoeff^x_{ij} = \Jcoeff^x_{ii} \Gmatcoeff^x_{ij} ,
\end{equation}
\noindent and $\Gcompcoeff^x_{ij} = [ \Gcomp_x ]_{i,j}$,
$\Gmatcoeff^x_{ij} = [ \Gmat_x ]_{i,j}$ and $\Jcoeff^x_{ij} = [
  \Jmat_x ]_{i,j}$ are the coefficients of the matrices $\Gcomp_x$,
$\Gmat_x$ and $\Jmat_x$, respectively. Since $\Gmat$ (also $\Gcomp$)
is usually a zero-diagonal matrix, \ie~$\Gmatcoeff_{ii} = 0$
(summation not implied), the condition {$\| \Gcomp \|_2 \le 1$}
simplifies to
\begin{equation}
\label{ineq_eigenG}
|\lambda_{i}^{\Gcomp_x} | \le \sum_{j \neq i} | \Gcompcoeff^x_{ij} | \le 1 \hspace{5mm} \forall i = 1, \dots, \pvlength ,
\end{equation}
\noindent where $\pvlength$ is the number of unknowns in our
domain. Finally, recalling that the Jacobian must be positive,
$\Jcoeff_{ii} > 0$, and extending the previous analysis to the $y$ and
$z$ directions, the following definition for the Jacobian
\begin{equation}
\label{unstrJacob}
\Jcoeff^x_{ii} \equiv \frac{1}{\sum_{j \neq i} | \Gmatcoeff^x_{ij} |} \hspace{10mm}
\Jcoeff^y_{ii} \equiv \frac{1}{\sum_{j \neq i} | \Gmatcoeff^y_{ij} |} \hspace{10mm}
\Jcoeff^z_{ii} \equiv \frac{1}{\sum_{j \neq i} | \Gmatcoeff^z_{ij} |} ,
\end{equation}
\noindent guarantees that inequalities~(\ref{ineq_eigenG}) are always
satisfied. {Here, no summation over $i$ is implied.} Therefore,
the spectral norm of $\Gmat_\xi$ is equal to or smaller than unity,
\ie~$\| \Gcomp \|_2 \le 1$. In this way, the local Jacobian for the
node $i$, $\J_i$, is given by
\begin{equation}
\label{locJacob}
\J_i = \left( \begin{array}{ccc} \Jcoeff^x_{ii} & & \\ & \Jcoeff^y_{ii} & \\ & & \Jcoeff^z_{ii} \end{array} \right) .
\end{equation}
{\noindent Notice that the definitions of the Jacobian given in
  Eq.~(\ref{unstrJacob}) are solely based on the coefficients of the
  discrete gradient operator, $\Gmat$. Therefore, there is no
  restriction regarding the type of grid and the numerical method.}
{Moreover, it} is worth noticing that for a Cartesian uniform
mesh, this formula reduces to $\J = \diag ( \Dx, \Dy, \Dz )$ similar to the
definition of $\DelTen$ given in Eq.~(\ref{DelTen_def}).

In this way, the {subgrid characteristic length scale} proposed in
Section~\ref{building} is straightforwardly extended to unstructured
meshes by simply replacing $\DelTen$ in Eq.~(\ref{DeltaLsq}) by the
local Jacobian, $\J_i$, defined in Eqs.~(\ref{unstrJacob})
and~(\ref{locJacob}).


\section{Numerical results}

\label{results}

\subsection{Decaying homogeneous isotropic turbulence}

\label{HIT}

\begin{figure}[!t]
\centering{
  \includegraphics[angle=-90,width=0.63\textwidth]{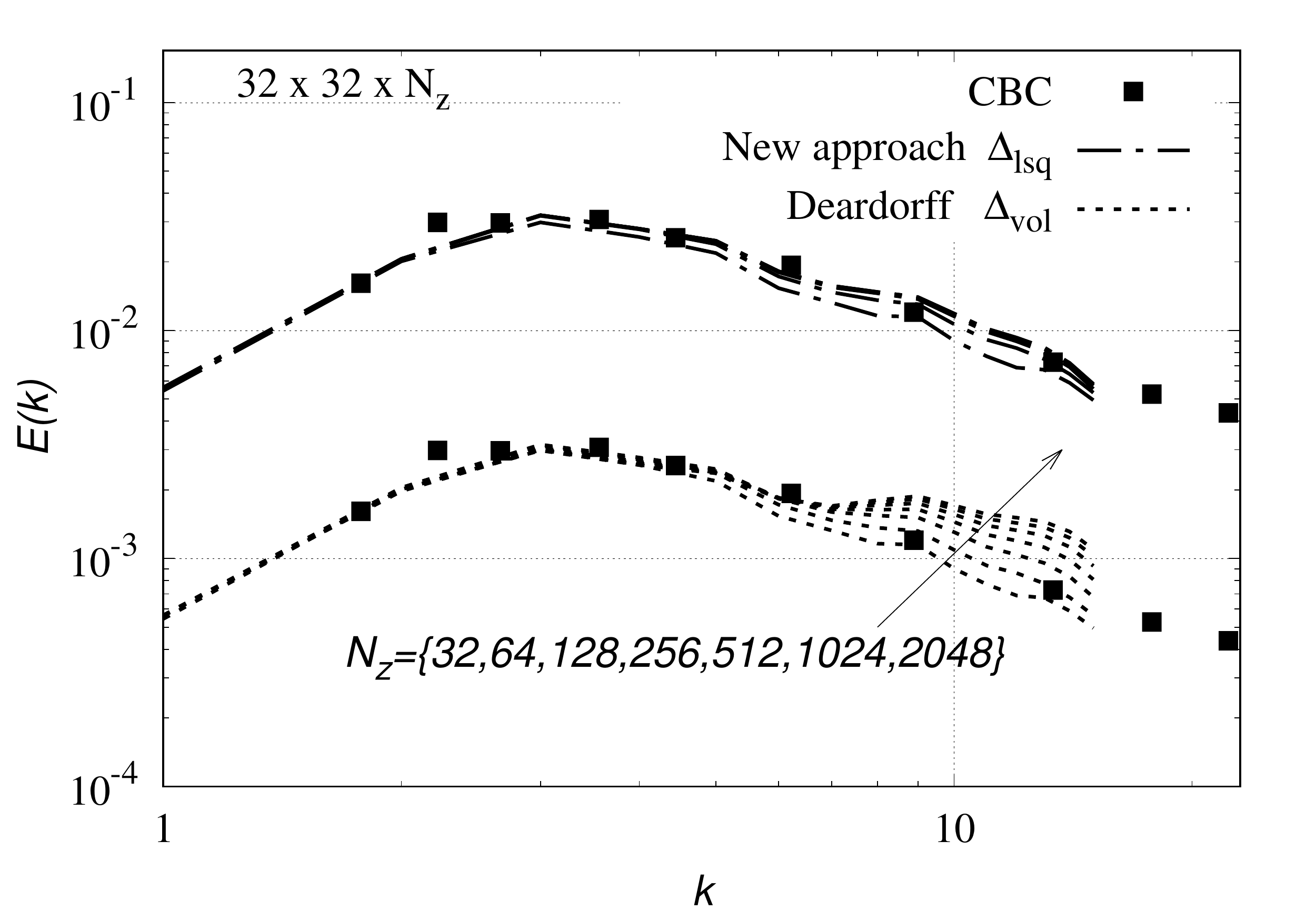}
  \includegraphics[angle=-90,width=0.63\textwidth]{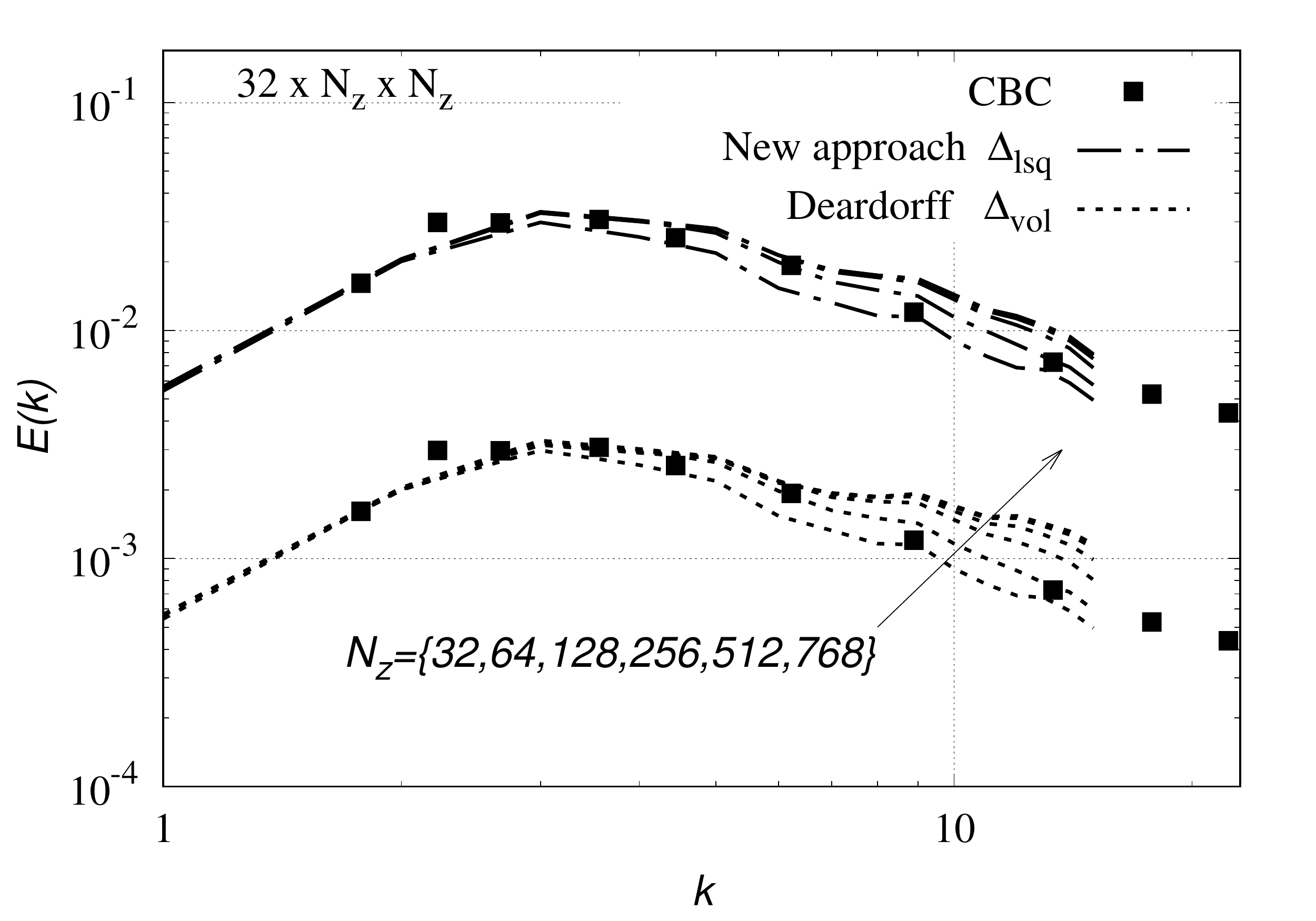}
}
\caption{Three-dimensional kinetic energy spectra as a function of
  computational wavenumber, for decaying isotropic turbulence
  corresponding to the experiment of Comte-Bellot and
  Corrsin~\cite{COM71}. LES results have been obtained using the
  Smagorinsky model for a set of anisotropic meshes with pancake-like
  (top) and pencil-like (bottom) control volumes. Results obtained
  with the novel definition of $\FLlsq$ proposed
  in~Eq.~(\ref{DeltaLsq}) are compared with the classical definition
  proposed by Deardorff given in Eq.~(\ref{DeltaDeardorff}). For
  clarity, the latter results are shifted one decade down.}
\label{CBCresults}
\end{figure}

The numerical simulation of decaying isotropic turbulence {was}
chosen as a first case to test the novel definition of {the
  subgrid characteristic length scale,} $\FLlsq${,} proposed
in~Eq.~(\ref{DeltaLsq}). The configuration corresponds to the classical
experiment of Comte-Bellot and Corrsin (CBC)~\cite{COM71}. Large-eddy
simulation results have been obtained using the Smagorinsky model, for
a set of (artificially) stretched meshes. Namely, results for
pancake-like meshes with $32 \times 32 \times \Nz$ and
$\Nz=\{32,64,128,256,512,1024,2048\}$ are displayed in
Figure~\ref{CBCresults} (top). As expected, for increasing values of
$\Nz$, the results obtained using the classical definition of
Deardorff, given in Eq.~(\ref{DeltaDeardorff}), diverge. This is
because the value of $\FLvol$ tends {to zero for increasing $\Nz$}
and, therefore, the subgrid-scale model switches off. This is not the
case for the definition of $\FLlsq$ proposed in this
work. Interestingly, the results rapidly converge for increasing
values of $\Nz$. Therefore, the proposed definition of the subgrid
characteristic length, $\FLlsq$, seems to minimize the effect of mesh
anisotropies on the performance of subgrid-scale models.

\begin{figure}[!t]
	\centering{
		\includegraphics[angle=-90,width=0.63\textwidth]{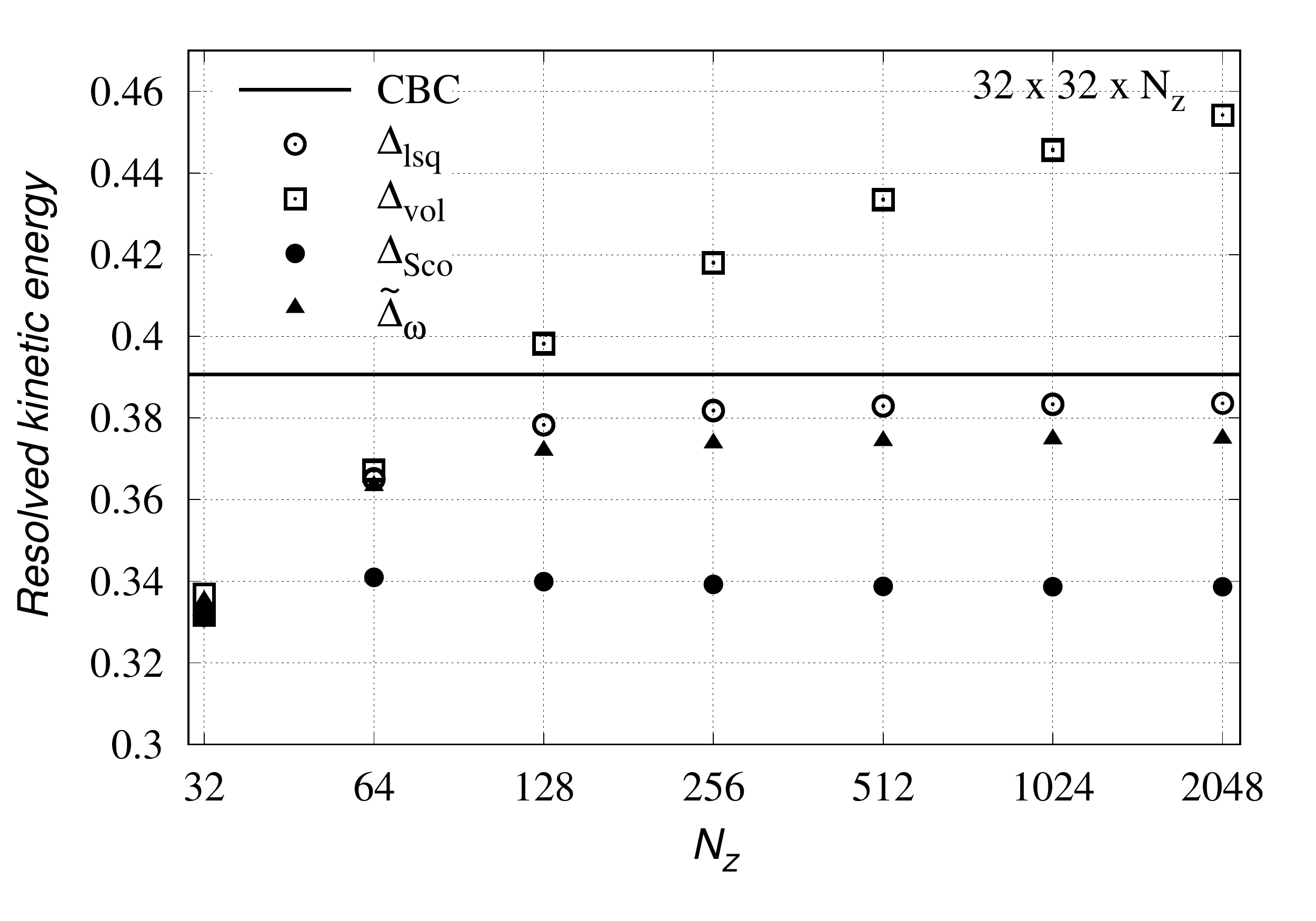}
		\includegraphics[angle=-90,width=0.63\textwidth]{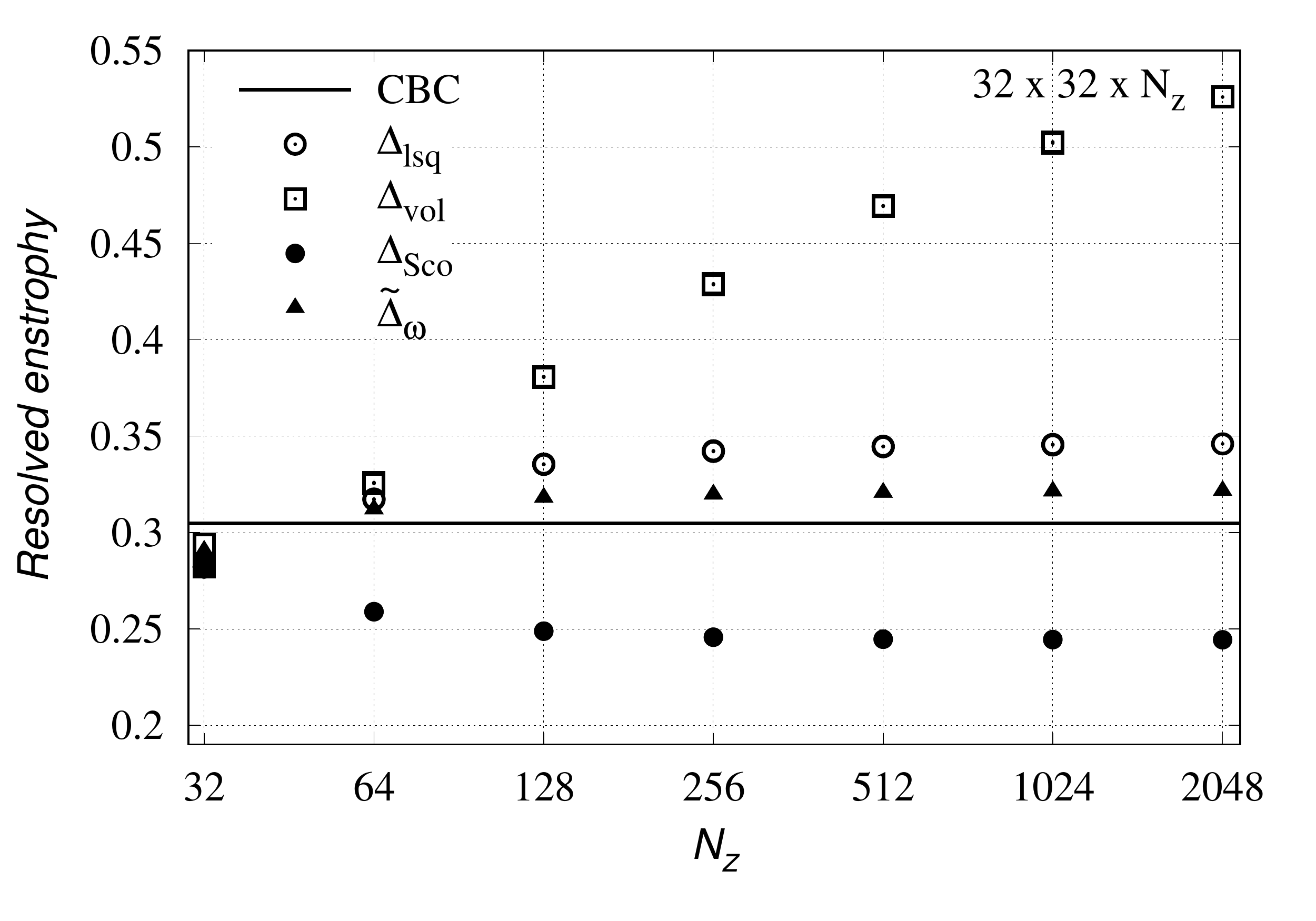}
	}
	\caption{{Resolved kinetic energy (top) and enstrophy (bottom) as
			a function of the number of grid points, $\Nz$, for decaying
			isotropic turbulence corresponding to the experiment of
			Comte-Bellot and Corrsin~\cite{COM71}. LES results have been
			obtained using the Smagorinsky model for a set of anisotropic
			meshes with pancake-like control volumes, \ie~$32 \times 32 \times
			\Nz$. Results obtained with the novel definition of $\FLlsq$
			proposed in Eq.~(\ref{DeltaLsq}) are compared with the definitions
			proposed by Deardorff~\cite{DEA70}, $\FLvol$
			[Eq.~(\ref{DeltaDeardorff})], Scotti~\etal~\cite{SCO93}, $\FLSco$
			[Eq.~(\ref{DeltaScotti})], and Mockett~\etal~\cite{MOC15}, $\FLMoc$
			[Eq.~(\ref{DeltaShu})], respectively.}}
	\label{CBCconvergence}
\end{figure}

Similar behavior is observed in Figure~\ref{CBCresults} (bottom) for
pencil-like meshes with $32 \times \Nz \times \Nz$ grid cells, where
$\Nz=\{32,64,128,256,512,768\}$. In this case, the improper behavior
of Deardorff's definition is even more evident because $\nut$ in
Eq.~(\ref{eddyvis_template}) scales as ${\mathcal O}(\Dz^{4/3})$
instead of the ${\mathcal O}(\Dz^{2/3})$ scaling for the pancake-like
meshes. Therefore, the model switches off even more
rapidly. Furthermore, it is worth mentioning that in this case this
numerical artifact is visible for a wide range of wavenumbers, whereas
for pancake-like meshes, only the smallest resolved scales are affected
in a significant manner. On the other hand, LES results obtained with
$\FLlsq$ also tend to converge for increasing values of
$\Nz$. Nevertheless, compared with the results obtained with
pancake-like meshes, significant changes are observed for the first
three meshes, \ie~$\Nz=\{32,64,128\}$. This delay in the convergence
of the LES results may be attributed to the fact that more scales are
actually being solved in two spatial directions (instead of one for
the pancake-like meshes). Therefore, the role of the LES model is
lessened, and differences in the results can be probably attributed to
the natural convergence {for grid refinement}. In any case,
compared with the classical Deardorff approach, the {newly proposed
  subgrid characteristic length scale} strongly reduces the artificial
effects caused by mesh anisotropies while providing a natural
convergence.

{In order to analyze in more detail the effect of mesh
  anisotropies for different definitions of $\Flength$, two physical
  quantities of interest have been studied: namely, the resolved
  kinetic energy and the resolved enstrophy. Results displayed in
  Figure~\ref{CBCconvergence} have been obtained with the same
  pancake-like meshes (\ie~ $32 \times 32 \times \Nz$ grid cells, where
  $\Nz=\{32,64,128,256,512,1024,2048\}$). In this case, apart from the
  new definition, $\FLlsq$, and the Deardorff length scale, $\FLvol$,
  two additional definitions have also been tested: the definition
  proposed by Scotti~\etal~\cite{SCO93}, $\FLSco$, given in
  Eq.~(\ref{DeltaScotti}) and the definition proposed by
  Mockett~\etal~\cite{MOC15}, $\FLMoc$, given in
  Eq.~(\ref{DeltaShu}). It must be noted that for this comparison, we
  have chosen $\FLSco$ because among all the definitions reviewed in
  Section~\ref{intr} that solely depend on geometrical properties of
  the mesh (see property {\bfseries P3} in Table~\ref{properties_Delta}) this is
  the length scale that provides the best results. Regarding the flow-dependent
  definitions, we have chosen $\FLMoc$ because this definition was
  actually proposed as an improvement of the definition by
  Chauvet~\etal~\cite{CHA07} given in Eq.~(\ref{Deltavort}). The
  definition proposed by Shur~\etal~\cite{SHU15}, $\FLSLA$, given in
  Eq.~(\ref{DeltaSLA}) has not been considered here because it is just
  a modification of $\FLMoc$ specifically adapted to trigger the \KH
  instability in the initial part of shear layers.}

{As explained above, energy spectra obtained using Deardorff's length scale $\FLvol$ diverge for increasing values
  of $\Nz$ due to the fact that $\FLvol$ tends to zero and,
  therefore, the subgrid-scale model switches off. This effect becomes
  even more evident for the resolved enstrophy (see
  Figure~\ref{CBCconvergence}, bottom) since this lack of SGS
  dissipation mainly affects the smallest resolved scales. This
  physically improper behavior is strongly mitigated by other
  definitions of $\Flength$. Namely, the definition proposed by
  Scotti~\etal~\cite{SCO93}, $\FLSco$, displays the weakest dependence with
  respect to $\Nz$. As explained in Section~\ref{intr}, this
  definition of $\Flength$ was proposed as a correction of the
  Deardorff definition, $\FLvol$, for anisotropic meshes with the
  assumption of an isotropic turbulent regime. Therefore, it is not
  surprising that this definition behaves very robustly for a
  simulation of decaying homogeneous isotropic turbulence. On the
  other hand, we can observe that the novel definition, $\FLlsq$, and
  the definition proposed by Mockett~\etal~\cite{MOC15}, $\FLMoc$,
  display a very similar behavior. Results for both resolved kinetic
  energy and enstrophy rapidly converge for increasing values of
  $\Nz$. Even more interestingly, taking the CBC results as an
  indication of the trend the data should have, both definitions lead
  to significantly better solutions compared with the original $32^3$
  mesh and the solution obtained with the definition proposed by
  Scotti~\etal~\cite{SCO93}, $\FLSco$.}

\begin{figure}
\centering{
  \includegraphics[angle=-90,width=0.63\textwidth]{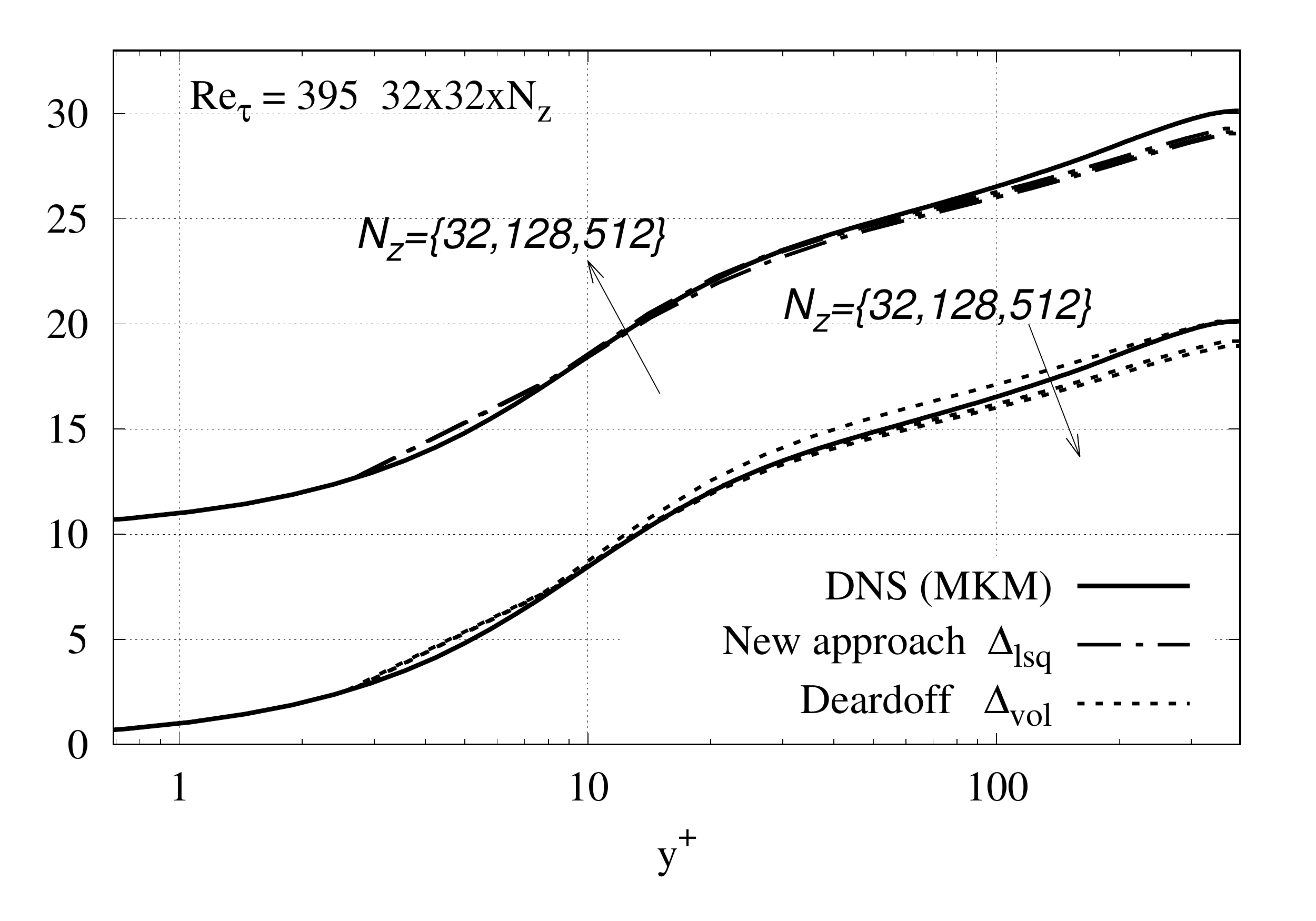}
  \includegraphics[angle=-90,width=0.63\textwidth]{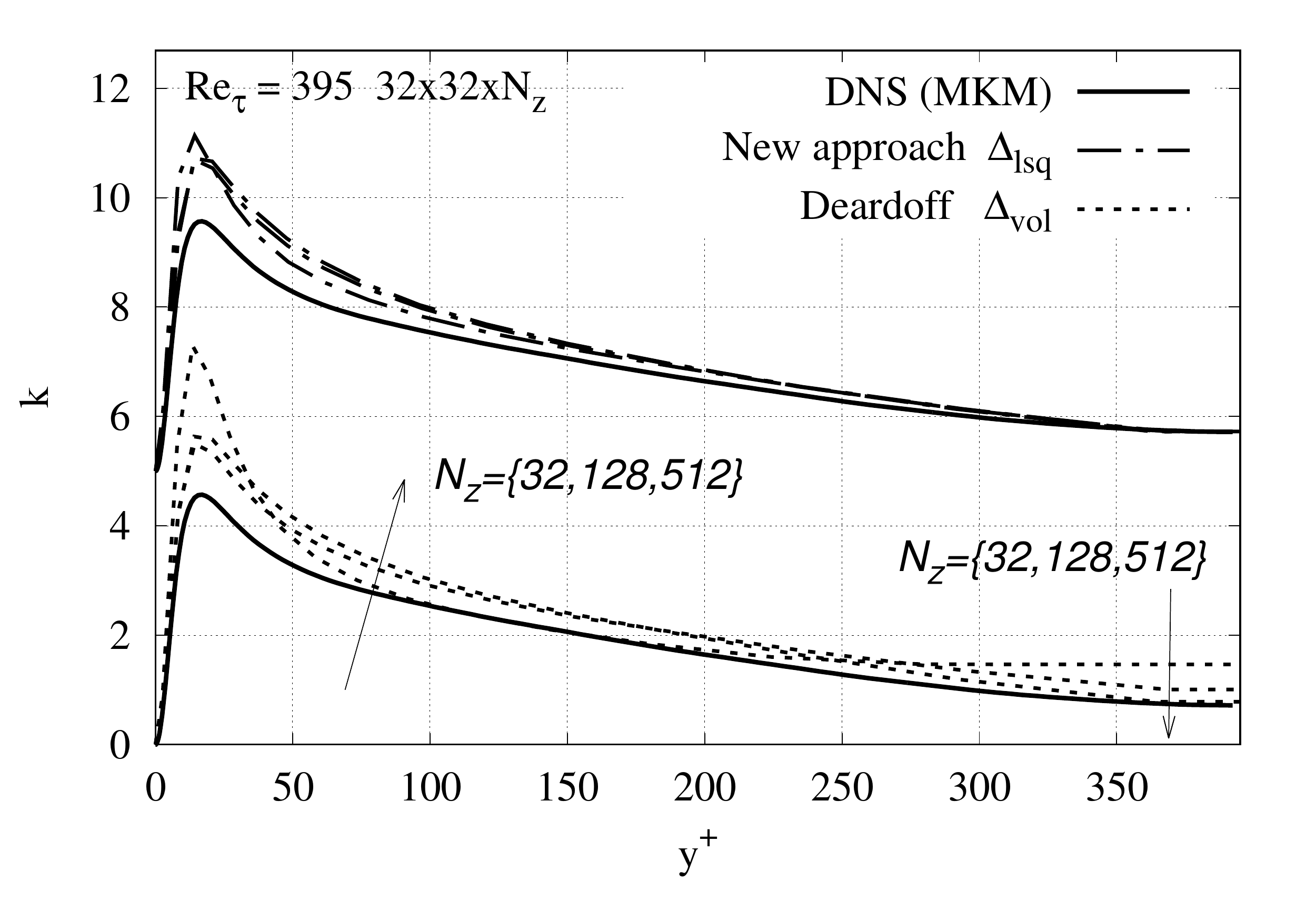}
}
\caption{Results for a turbulent channel flow at $Re_{\tau}=395$
  obtained with a set of anisotropic meshes using the S3PQ
  model~\cite{TRI14-Rbased}. Solid lines correspond to the direct
  numerical simulation of Moser~\etal~\cite{MOS99}. Results obtained
  with the novel definition of $\FLlsq$ proposed in
  Eq.~(\ref{DeltaLsq}) are compared with the classical definition
  proposed by Deardorff given in Eq.~(\ref{DeltaDeardorff}). For
  clarity, the former results are shifted up. Top: mean streamwise
  velocity, $\avgtime{u}$. Bottom: turbulent kinetic energy, $k$.}
\label{results_CF}
\end{figure}

\begin{figure}
	\centering{
		\includegraphics[angle=-90,width=0.63\textwidth]{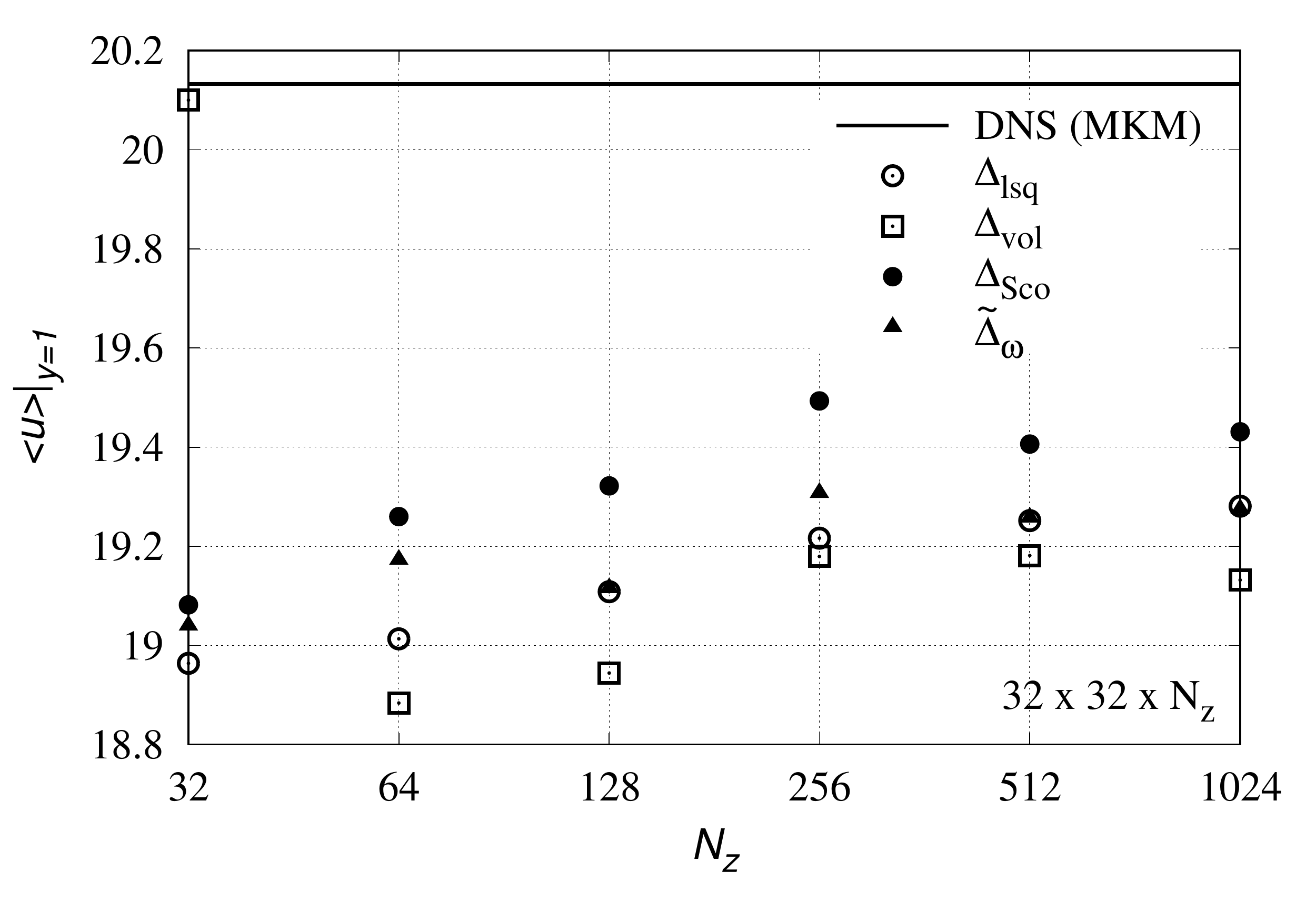}
		\includegraphics[angle=-90,width=0.63\textwidth]{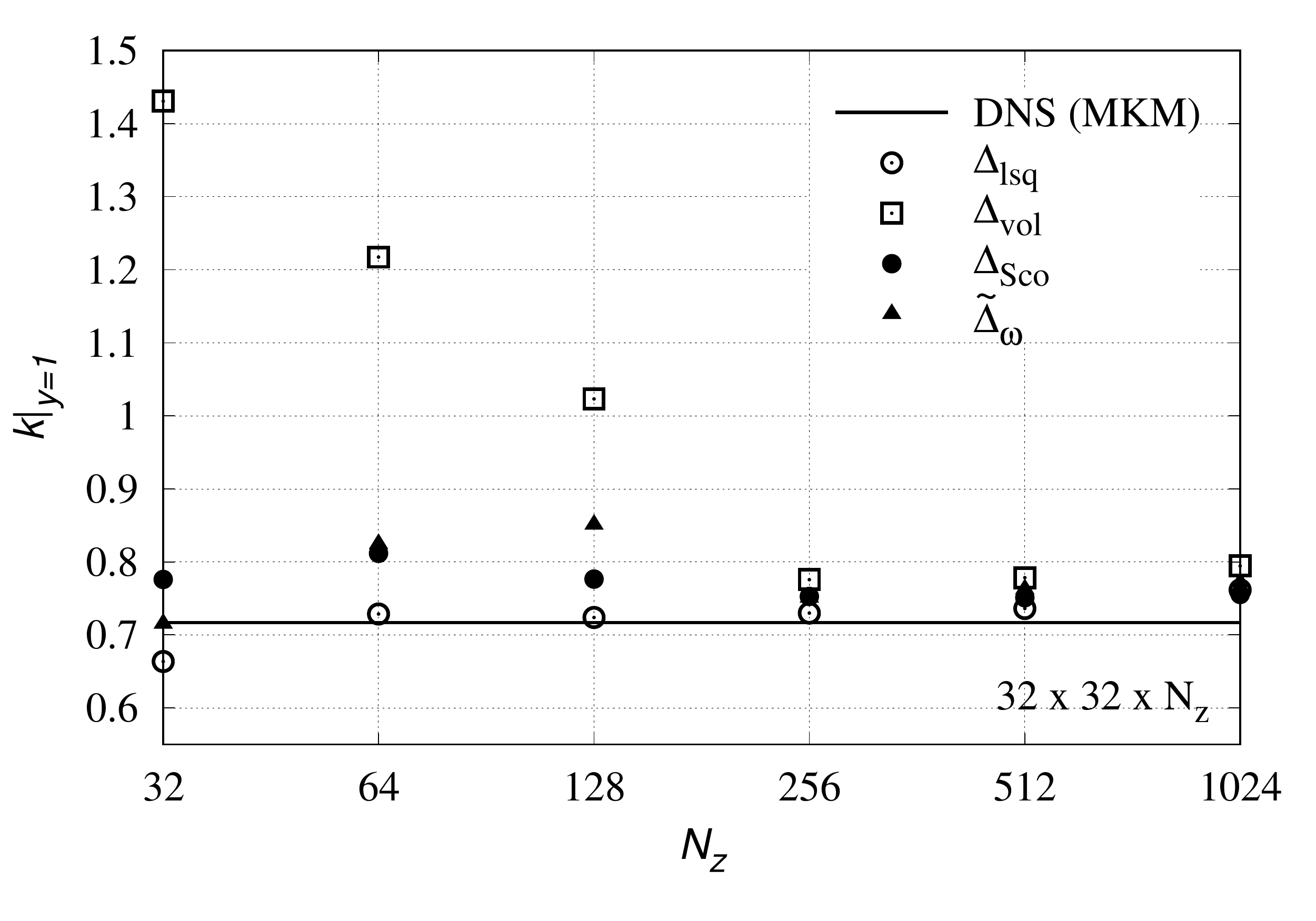}
	}
	\caption{{Results for a turbulent channel flow at $Re_{\tau}=395$
			obtained with a set of anisotropic meshes using the S3PQ
			model~\cite{TRI14-Rbased}. Solid lines correspond to the direct
			numerical simulation of Moser~\etal~\cite{MOS99}. Results obtained
			with the novel definition of $\FLlsq$ proposed in
			Eq.~(\ref{DeltaLsq}) are compared with the definitions proposed by
			Deardorff~\cite{DEA70}, $\FLvol$ [Eq.~(\ref{DeltaDeardorff})],
			Scotti~\etal~\cite{SCO93}, $\FLSco$ [Eq.~(\ref{DeltaScotti})] and
			Mockett~\etal~\cite{MOC15}, $\FLMoc$ [Eq.~(\ref{DeltaShu})],
			respectively. Top: mean streamwise velocity at channel
			mid-height, $\left.\avgtime{\uvel}\right|_{\y=1}$. Bottom:
			turbulent kinetic energy at channel mid-height,
			$\left.k\right|_{\y=1}$.}}
	\label{CFconvergence}
\end{figure}

\subsection{Turbulent channel flow}

\label{channel}

To test the performance of the proposed definition of $\FLlsq$ with
the presence of walls, simulations of a turbulent channel flow have
also been considered. In this case, the code is based on a
fourth-order symmetry-preserving finite-volume
discretization~\cite{VER03} of the incompressible Navier--Stokes
equations on structured staggered grids. Regarding the spatial
discretization of the eddy-viscosity models, the approach proposed by
Trias~\etal~\cite{TRI12-DmuS} has been used in conjunction with the
S3QR model recently proposed by
Trias~\etal~\cite{TRI14-Rbased}. Namely,
\begin{equation}
\label{S3QR} 
\nut^{S3QR} =  ( C_{s3qr} \Flength )^2 \Q_{\G\G\traspose}^{-1} \R_{\G\G\traspose}^{5/6} ,
\end{equation}
\noindent where $C_{s3pq} = 0.762$, $\Q_{\G\G\traspose}$ and
$\R_{\G\G\traspose}$ are the second and third invariants of the
symmetric second-order tensor $\G\G\traspose$, and $\G$ is the gradient
of the resolved velocity field, \ie~$\G \equiv \nabla \F{\vel}$. Similar to
Vreman's model~\cite{VRE04b}, the S3QR model is also based on the
invariants of the second-order tensor $\G\G\traspose$. However, it
{was designed to have} the proper cubic near-wall behavior. Apart
from this, it fulfills a set of desirable properties, namely,
positiveness, locality, Galilean invariance, and it automatically
switches off for laminar, 2D and axisymmetric flows. Furthermore, it
is well conditioned, has a low computational cost and has no intrinsic
limitations for statistically inhomogeneous flows.

Figure~\ref{results_CF} shows the results obtained from numerical
simulations of a turbulent channel flow at $Re_\tau = 395$ for a set
of (artificially) {refined} grids. The results are compared with
the DNS data of Moser~\etal~\cite{MOS99}. The dimensions of the
channel are taken equal to those of the DNS, \ie~$2\pi \times 2 \times
\pi$. The starting point corresponds to a $32^3$ mesh, which suffices
to obtain a good agreement with the DNS data. Therefore, the
computational grid is very coarse in comparison with the DNS which was
performed on a $256 \times 193 \times 192$ grid, \ie~the DNS used
about $290$ times more grid points than this first simulation. The
grid points are uniformly distributed in the stream-wise and the
span-wise directions, whereas the wall-normal points are distributed
using hyperbolic sine functions. For the lower half of the channel the
distribution of points is given by
\begin{equation}
y_j = \sinh ( \gamma j / \Ny ) / \sinh ( \gamma / 2 )  \hspace{5mm} j=0,1,..., \Ny/2 ,
\end{equation}
\noindent where $\Ny$ denotes the number of grid points in the
wall-normal direction. The stretching parameter, $\gamma$, is taken
equal to $7$. Then, the grid points in the upper half of the channel are computed by
means of symmetry. With this distribution and $\Ny = 32$, the first
off-wall grid point is located at $y^+ \approx 2.6$, \ie~inside the
viscous sublayer ($y^+ < 5$), {whereas $\Dx^+ \approx 77.5$ and
  $\Dz^+ \approx 38.8$. Hence, the grid is highly anisotropic in the
  near-wall region, \eg~$\Dx^+/\Dy^+ \approx 14.7$ for the first
  off-wall control volume.} 

Apart from the first simulation, two
additional meshes (with $N_z=128$ and $N_z=512$) have been used to
investigate the effect of $\Flength$. We chose to refine in the
span-wise direction because simulation results should not be too much
affected compared with the other two directions. Again, {as can be seen from Figure~\ref{results_CF}, }the results
obtained with the new definition of $\Flength$ are much more robust to
mesh anisotropies. It is remarkable that almost no changes are
observed in the mean velocity profile when the newly proposed length
scale is employed, whereas significant changes are observed for
Deardorff's classical definition. Similar behavior is observed for the
resolved turbulent kinetic energy, especially in the bulk region where
results obtained with the new length scale are almost independent of
the value of $N_z$.

{To study the effect of mesh anisotropies for other
  definitions of $\Flength$, results of the mean stream-wise velocity
  and turbulent kinetic energy at channel mid-height are displayed in
  Figure~\ref{CFconvergence}. Similar to the previous test case, results
  obtained using the definitions proposed by
  Scotti~\etal~\cite{SCO93}, $\FLSco$, given in Eq.~(\ref{DeltaScotti})
  and by Mockett~\etal~\cite{MOC15}, $\FLMoc$, given in
  Eq.~(\ref{DeltaShu}) are also shown for comparison. In this case,
  similar to the simulation of decaying homogeneous isotropic
  turbulence, the results obtained using Deardorff's definition,
  $\FLvol$, are strongly influenced by the mesh anisotropy. Again,
  other definitions tend to mitigate this. Despite the fact that it is based on the
  assumption of isotropic turbulence, the robustness
  of the definition proposed by Scotti~\etal~\cite{SCO93} regarding
  the turbulent kinetic energy is remarkable (see Figure~\ref{CFconvergence},
  bottom). However, its behavior is not so satisfactory regarding the
  average velocity field in the center of the channel (see
  Figure~\ref{CFconvergence}, top). The least to be expected from
  numerical simulations of turbulence is a robust prediction of the
  mean flow; therefore, the new definition, $\FLlsq$, and the
  definition $\FLMoc$ proposed by Mockett~\etal~\cite{MOC15}, display
  a significantly more robust behavior in this regard. However, the
  definition $\FLMoc$ is not so robust when predicting the turbulent
  kinetic energy (see Figure~\ref{CFconvergence}, bottom) where the
  results obtained with the new definition, $\FLlsq$ are almost not
  affected when $\Nz$ increases. In summary, the results obtained
  using the new length scale, $\FLlsq$, are at least as good as the best
  results obtained by other definitions with the advantage of having a
  much lower computational cost compared with $\FLMoc$ and being much
  easier to be used for unstructured grids.}


 \begin{figure}[!ht]
\centering{
  \includegraphics[angle=0,height=0.43\textwidth]{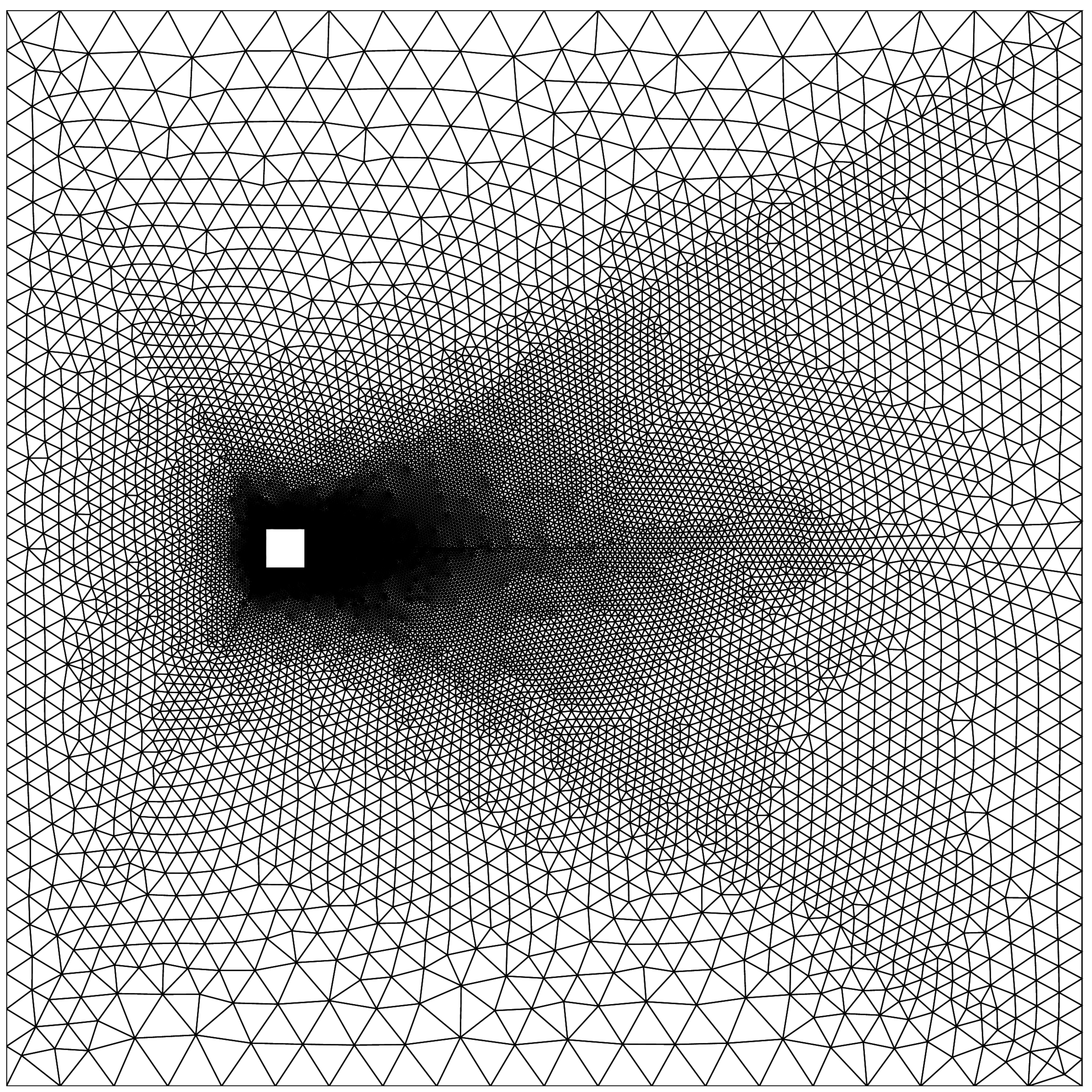}
  \hspace{3mm}
  \includegraphics[angle=0,height=0.43\textwidth]{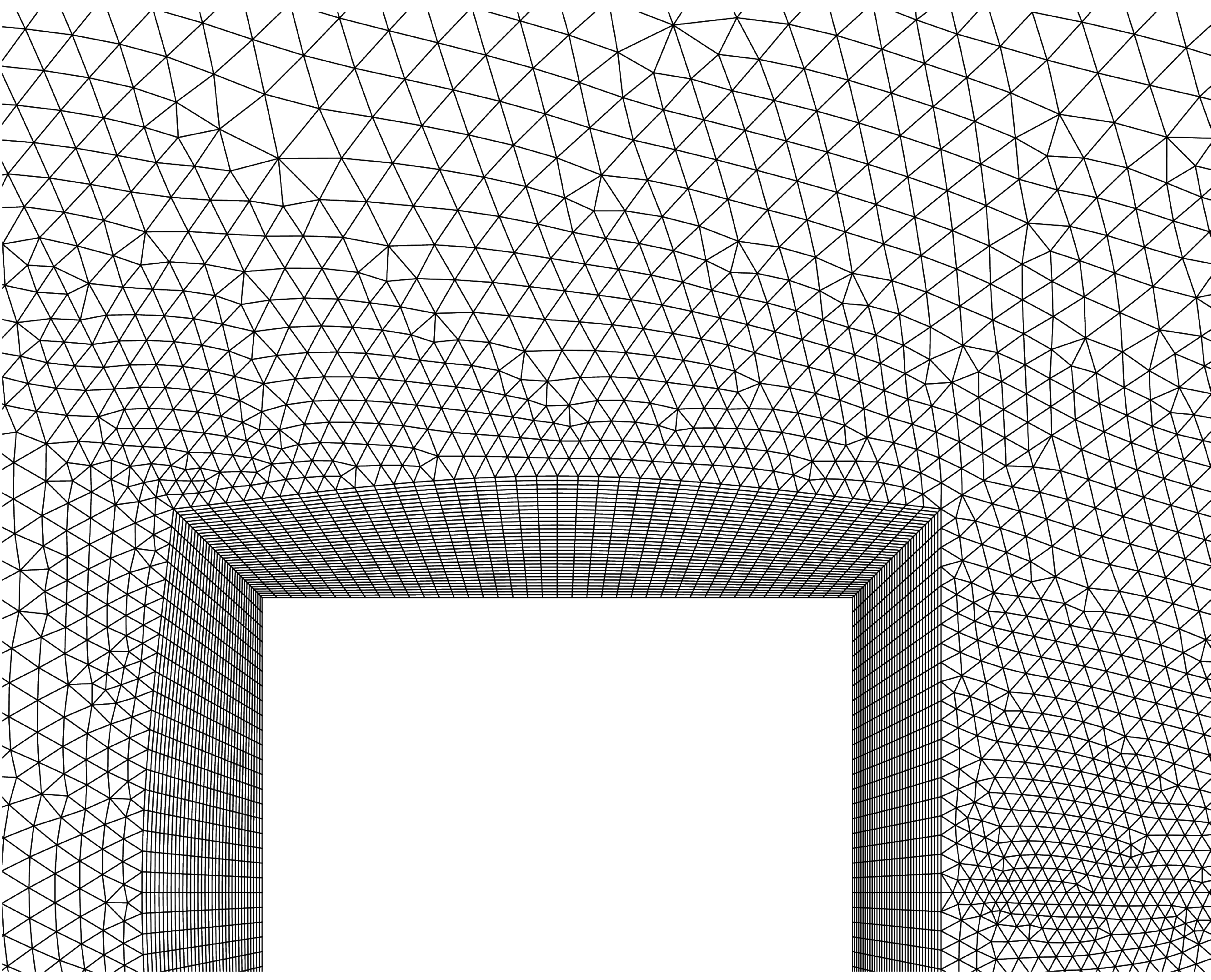}
  }
\caption{Left: 2D section of the unstructured mesh used to perform the
  set of LESs of the turbulent flow around a square
  cylinder at $Re=22000$. This $(\x,\y)$-section contains $19524$
  nodes. Right: zoom around the obstacle.}
\label{CYL22K_Mesh}
\end{figure}

\begin{figure}[!p]
\centering{
  \includegraphics[angle=0,width=0.43\textwidth]{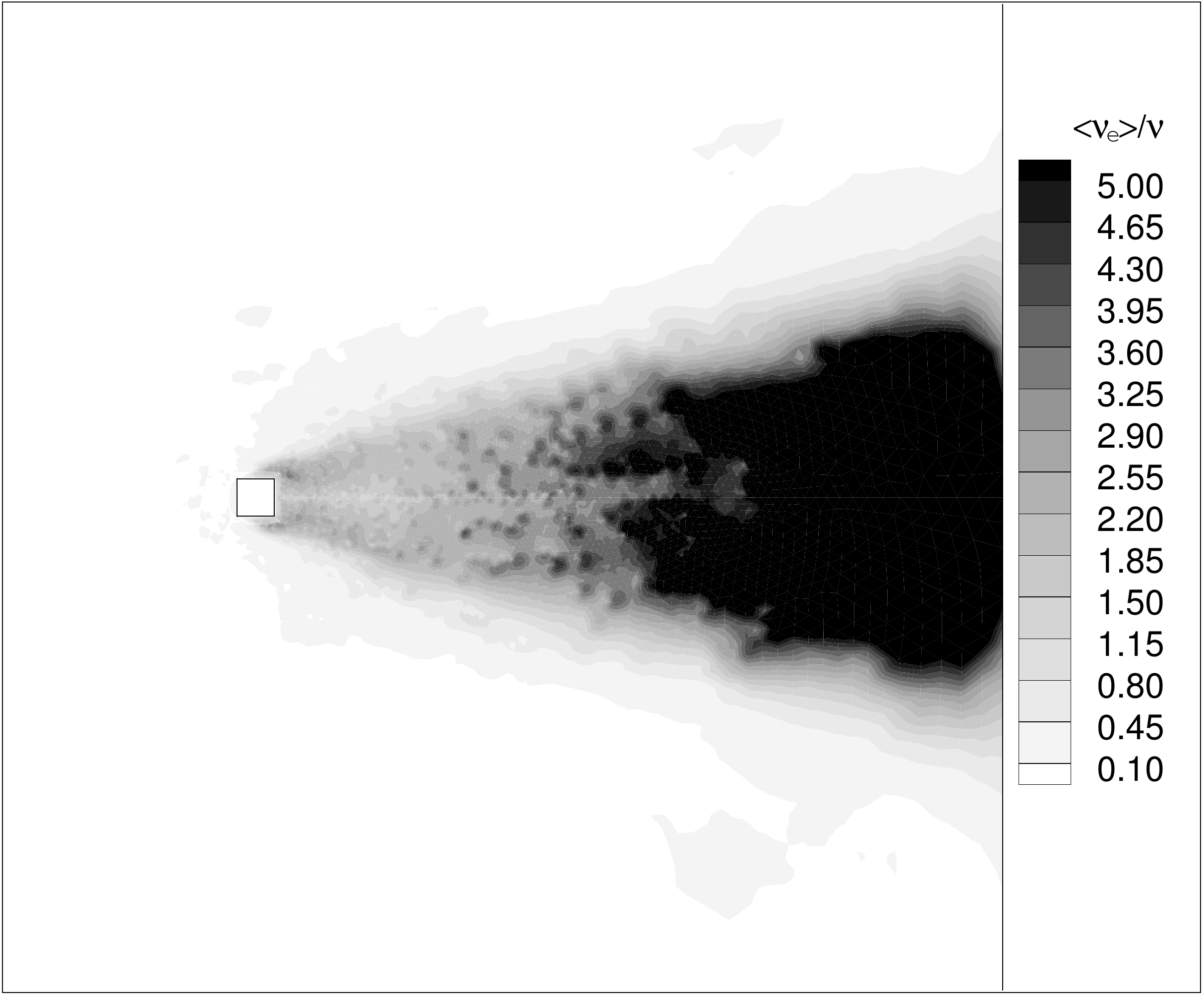}
  \hspace{3mm}
  \includegraphics[angle=0,width=0.43\textwidth]{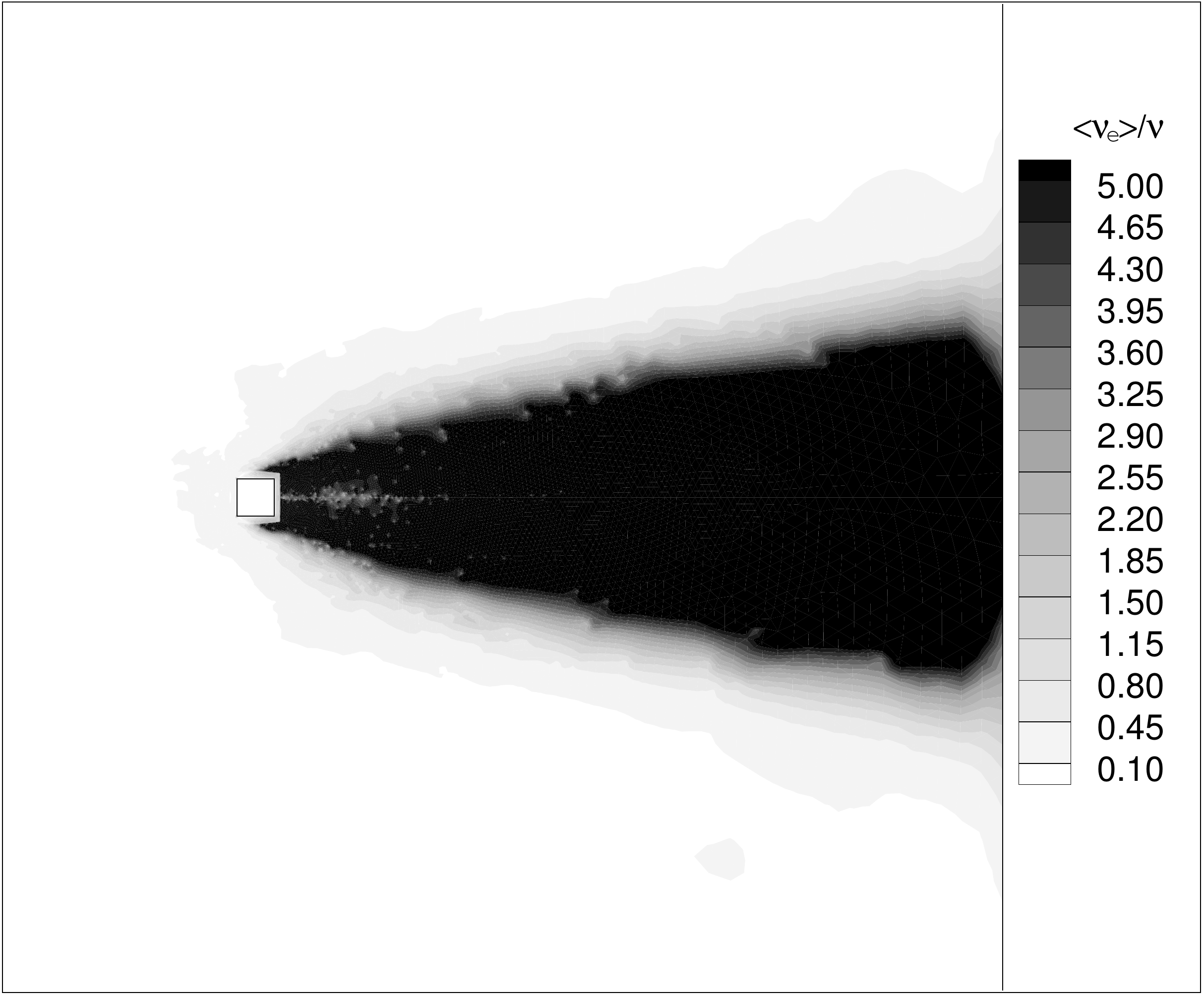}
}
\vspace{0.5mm}
\centering{
  \includegraphics[angle=0,width=0.43\textwidth]{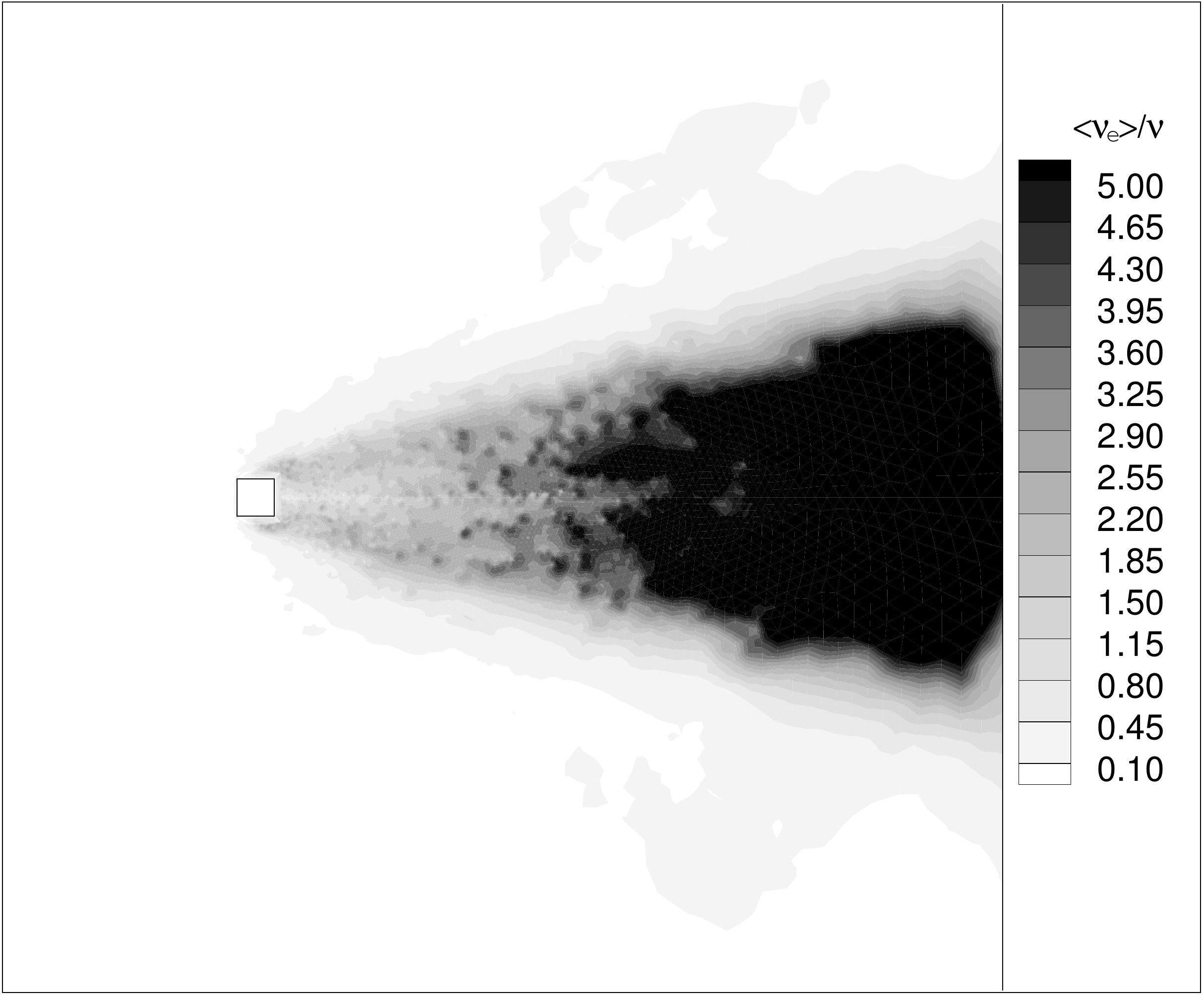}
  \hspace{3mm}
  \includegraphics[angle=0,width=0.43\textwidth]{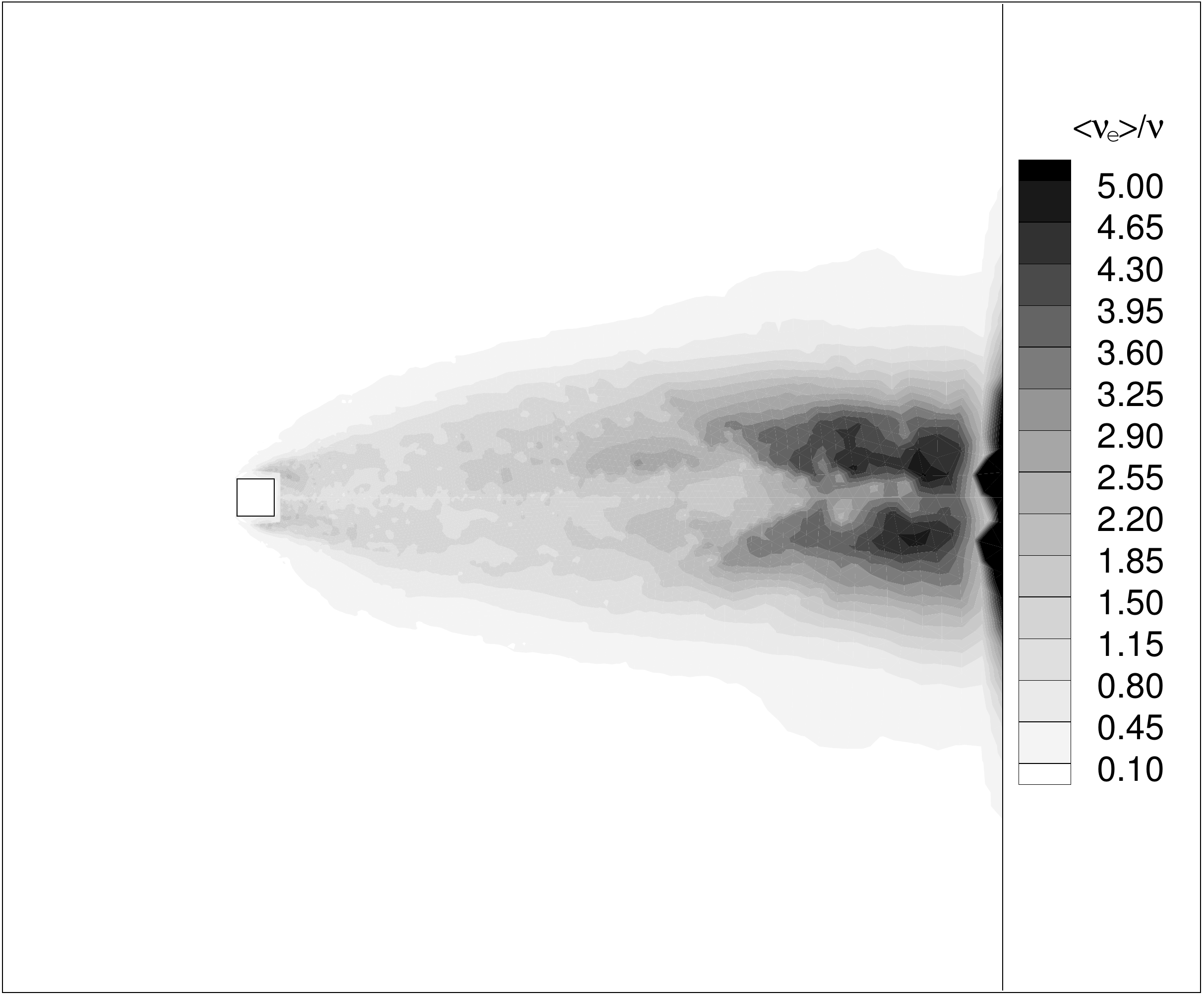}
  }
\caption{Average eddy-viscosity, $\avgtime{\nut}$, divided by the
  kinematic viscosity, $\nu$. Results for a turbulent flow around a
  square cylinder at $Re=22000$ carried out using the S3PQ
  model~\cite{TRI14-Rbased} on a set of unstructured meshes obtained
  from the extrusion in the span-wise direction of the 2D mesh
  displayed in Figure~\ref{CYL22K_Mesh}. Results obtained with the new
  definition, $\FLlsq$, proposed in Eq.~(\ref{DeltaLsq}) (left) are
  compared with the classical definition proposed by Deardorff given
  in Eq.~(\ref{DeltaDeardorff}) (right). This is done for two meshes
  differing in the number of grid points in the span-wise direction:
  $\Nz=100$ (top) and $\Nz=1000$ (bottom).}
\label{CYL22K_nut}
\end{figure}

\begin{figure}[!t]
\centering{
  \includegraphics[angle=0,width=0.43\textwidth]{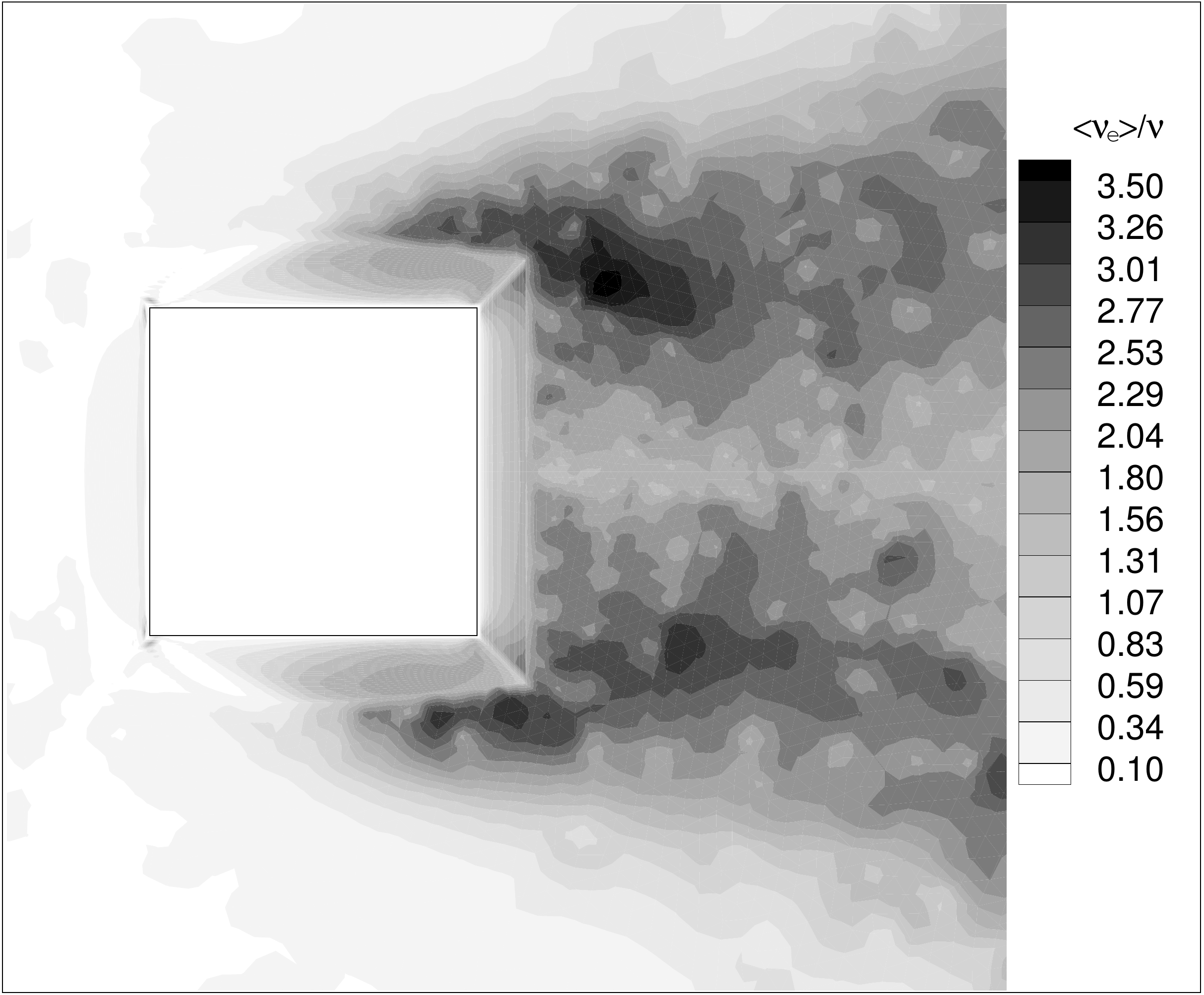}
  \hspace{3mm}
  \includegraphics[angle=0,width=0.43\textwidth]{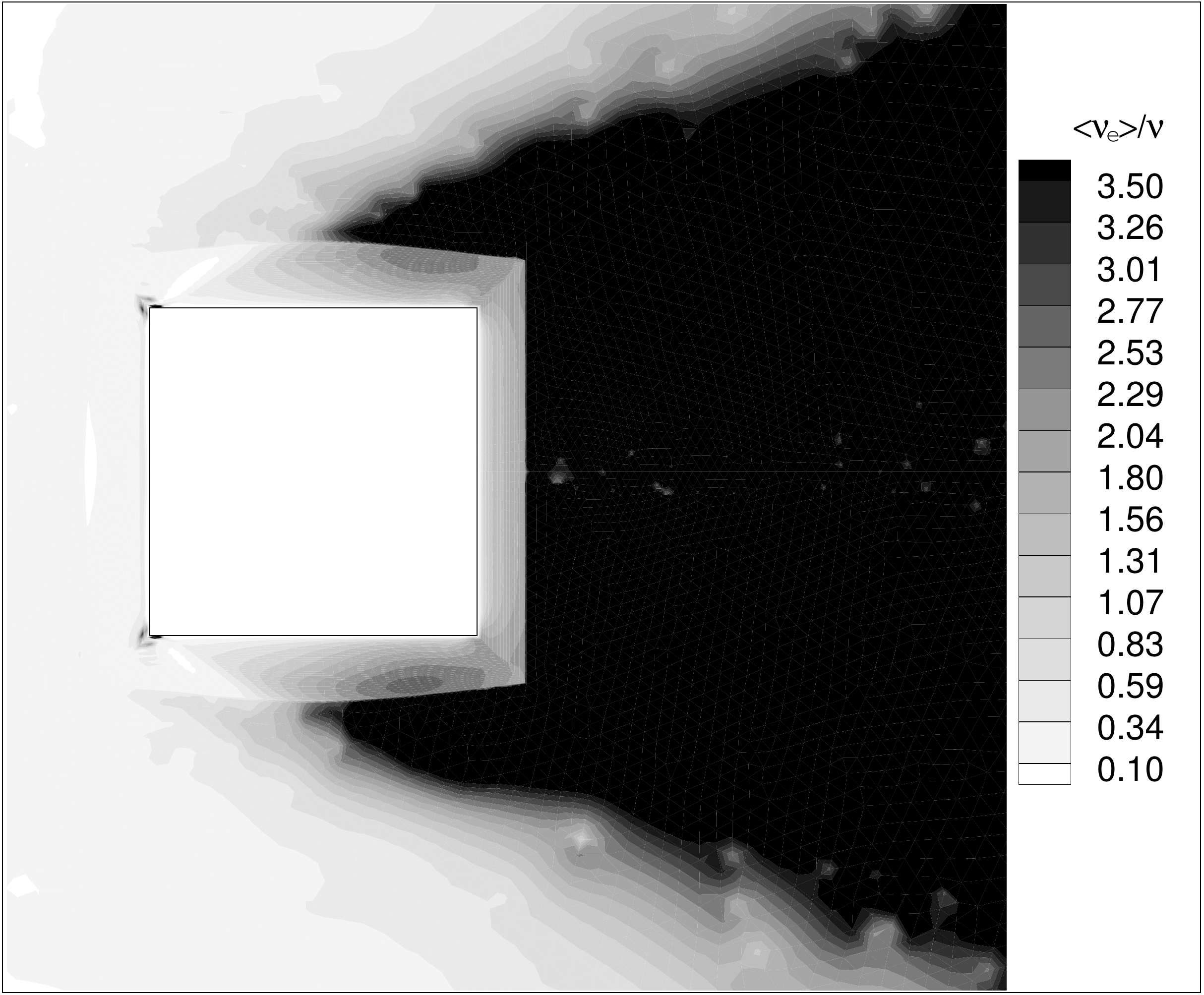}
}
\vspace{0.5mm}
\centering{
  \includegraphics[angle=0,width=0.43\textwidth]{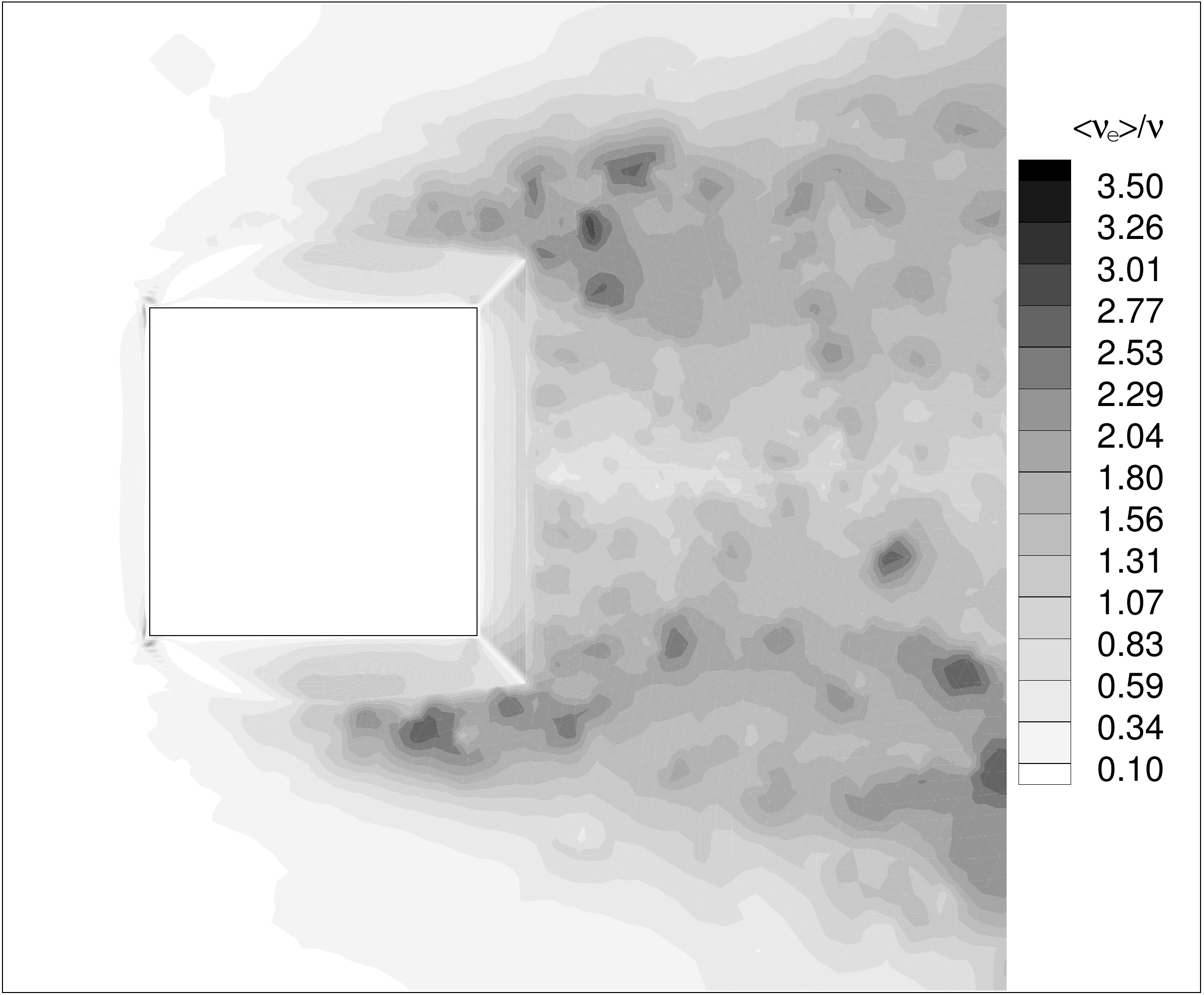}
  \hspace{3mm}
  \includegraphics[angle=0,width=0.43\textwidth]{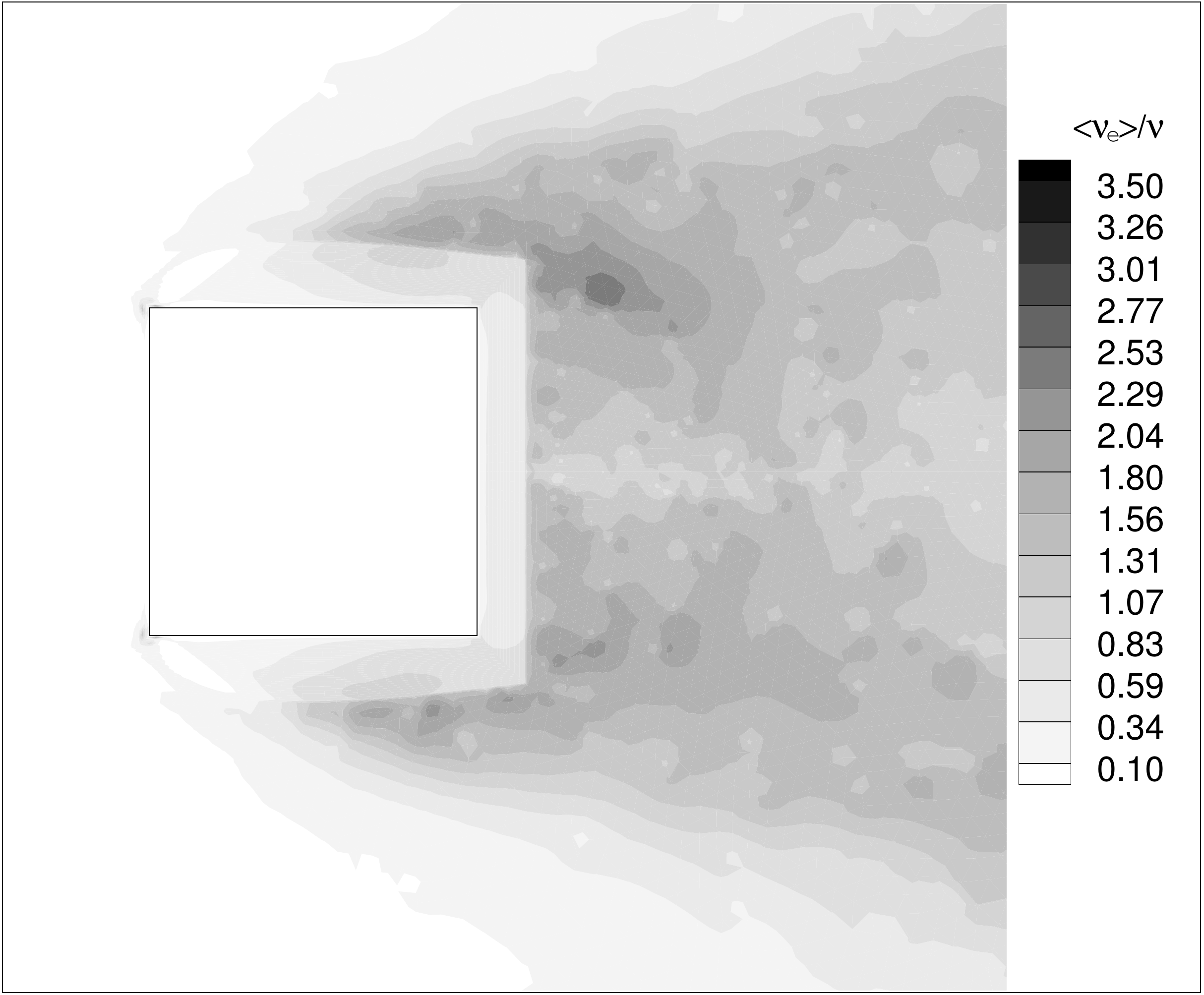}
  }
\caption{The same as in Figure~\ref{CYL22K_nut}. Zoom around the
  obstacle. Notice that the scale range has been properly modified.}
\label{CYL22K_nut_zoom2}
\end{figure}

\begin{figure}[!p]
\centering{
	\includegraphics[angle=-90,width=0.63\textwidth]{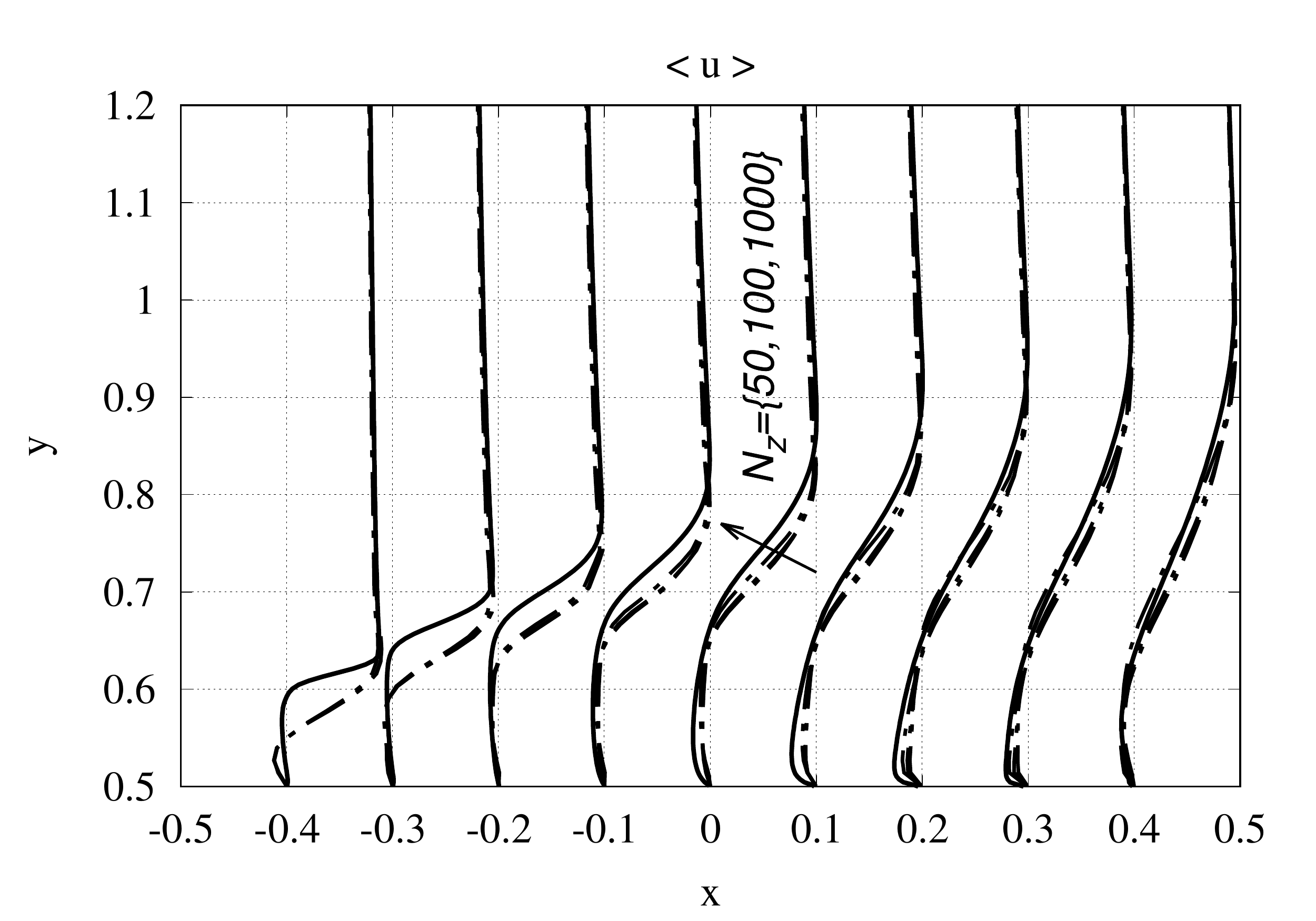}
	\includegraphics[angle=-90,width=0.63\textwidth]{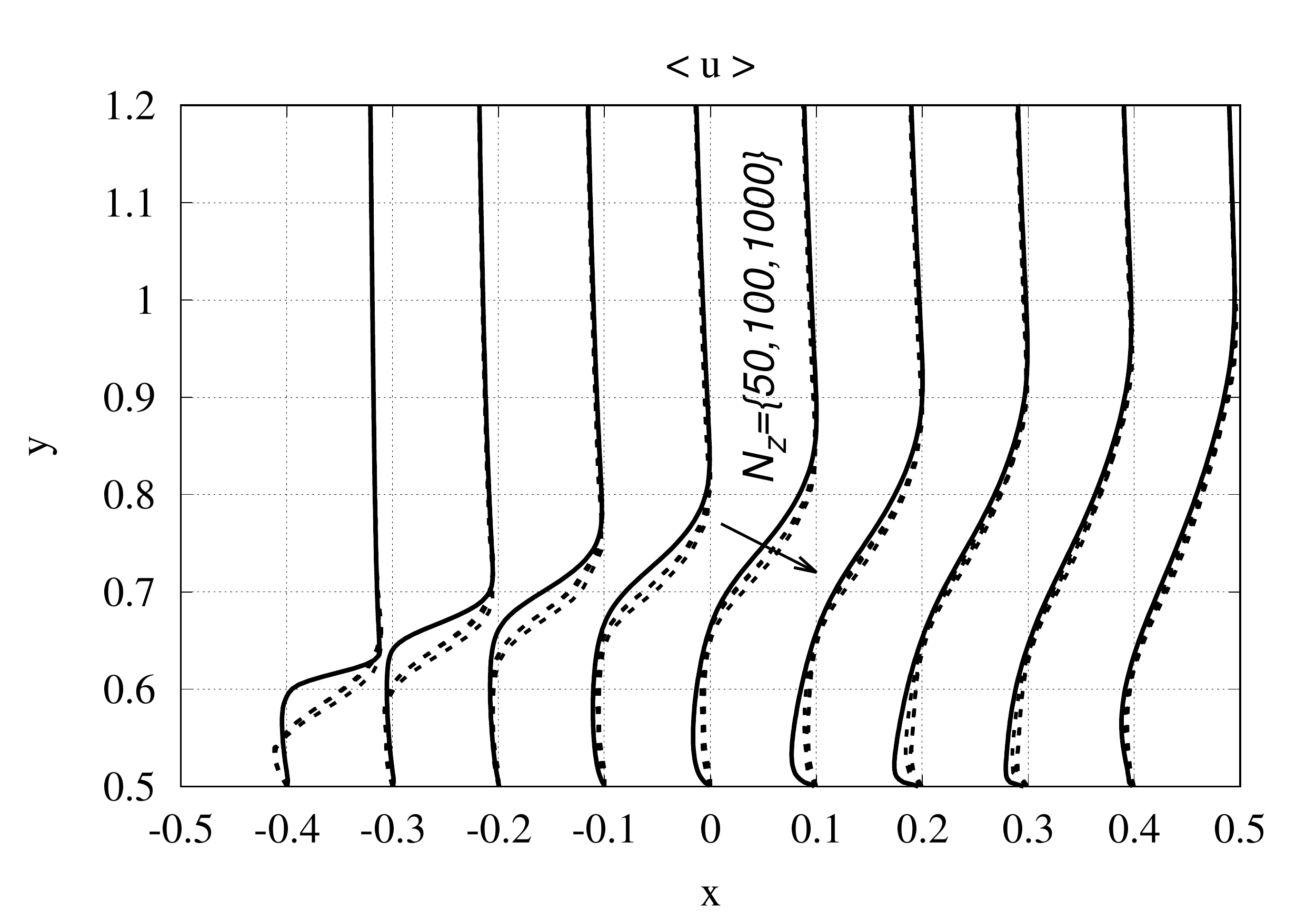}
}
\caption{Profiles of the average stream-wise velocity,
	$\avgtime{\uvel}$, in the near-wall region. Results correspond to a
	turbulent flow around a square cylinder at $Re=22000$ carried out
	using the S3PQ model~\cite{TRI14-Rbased} on a set of unstructured
	meshes obtained from the extrusion in the span-wise direction, \ie
	~$\Nz=\{50,100,1000\}$, of the 2D mesh displayed in
	Figure~\ref{CYL22K_Mesh}. Solid lines correspond to the DNS results
	of Trias~\etal~\cite{TRI14-CYL}. Results obtained with the novel
	definition $\FLlsq$ (top) proposed in Eq.~(\ref{DeltaLsq}) are
	compared with the classical definition proposed by Deardorff
	(bottom) given in Eq.~(\ref{DeltaDeardorff}).}
\label{CYL22K_U_BL}
\end{figure}

\begin{figure}[!t]
\centering{
	\includegraphics[angle=-90,width=0.63\textwidth]{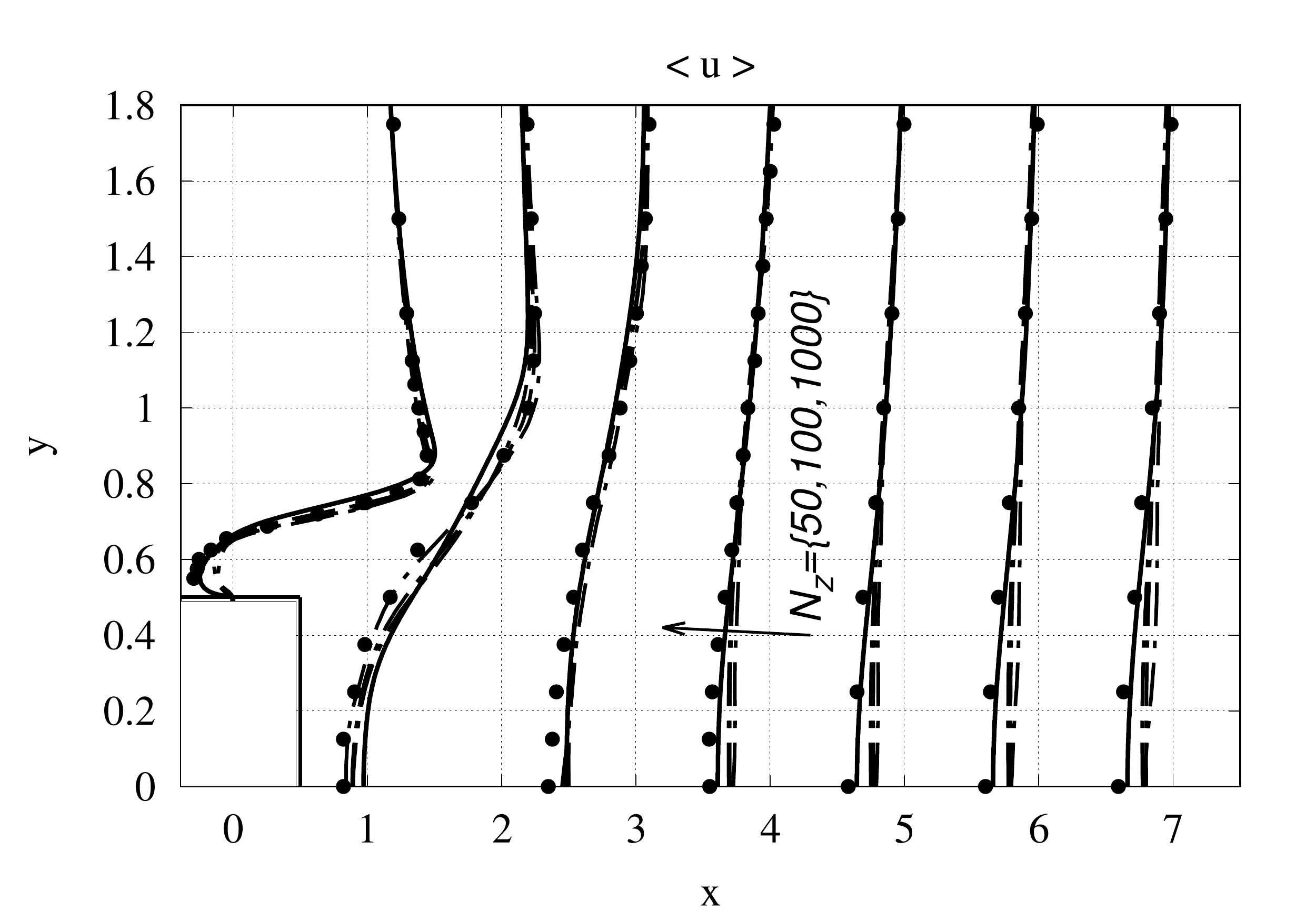}
	\includegraphics[angle=-90,width=0.63\textwidth]{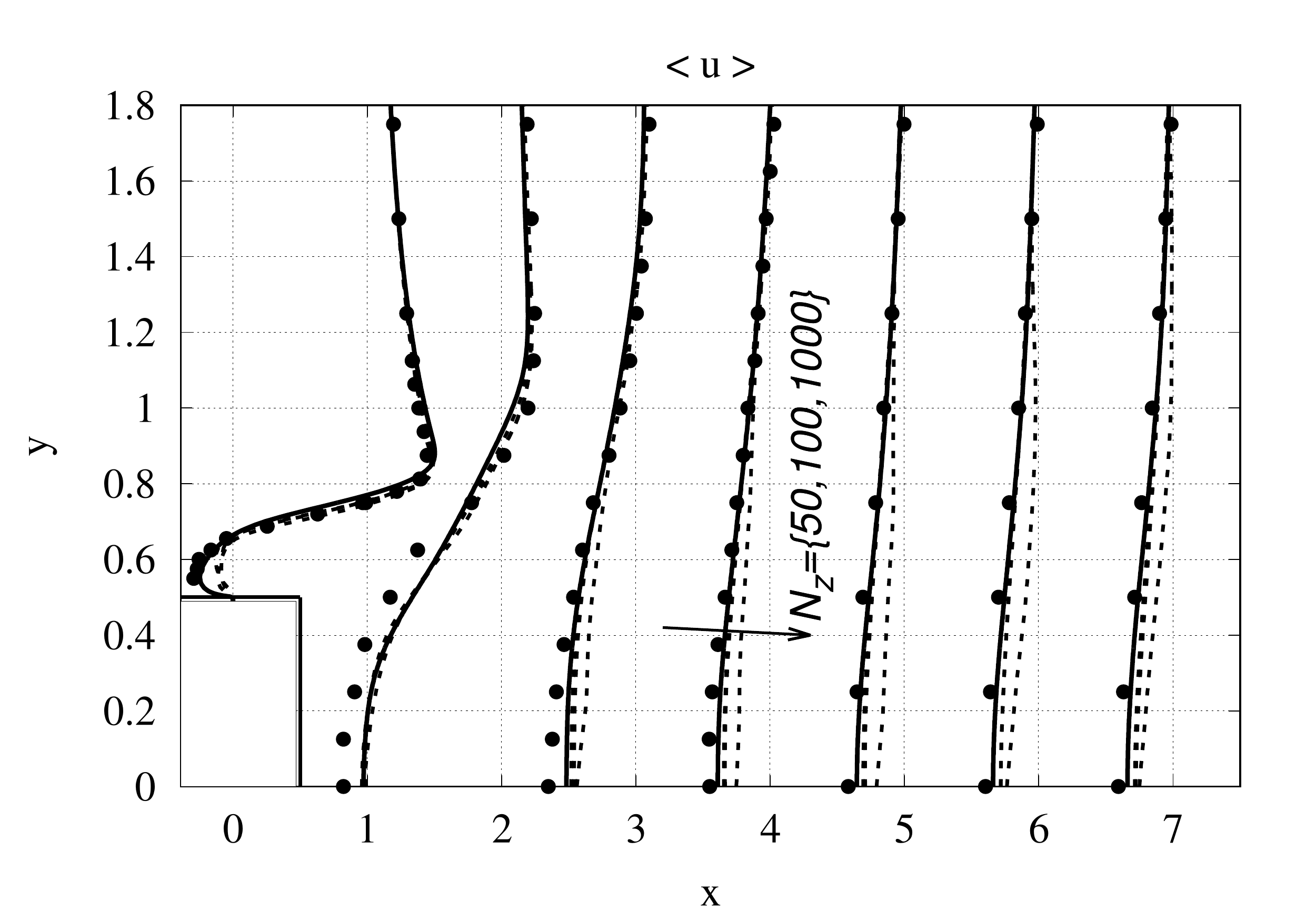}
}
\caption{The same as in Figure~\ref{CYL22K_U_BL}. In this case,
	profiles are in the wake region. Experimental results of
	Lyn~\etal~\cite{LYN95} (solid circles) are also displayed for
	comparison.}
\label{CYL22K_U_Wake}
\end{figure}

\begin{figure}[!t]
	\centering{
		\includegraphics[angle=-90,width=0.63\textwidth]{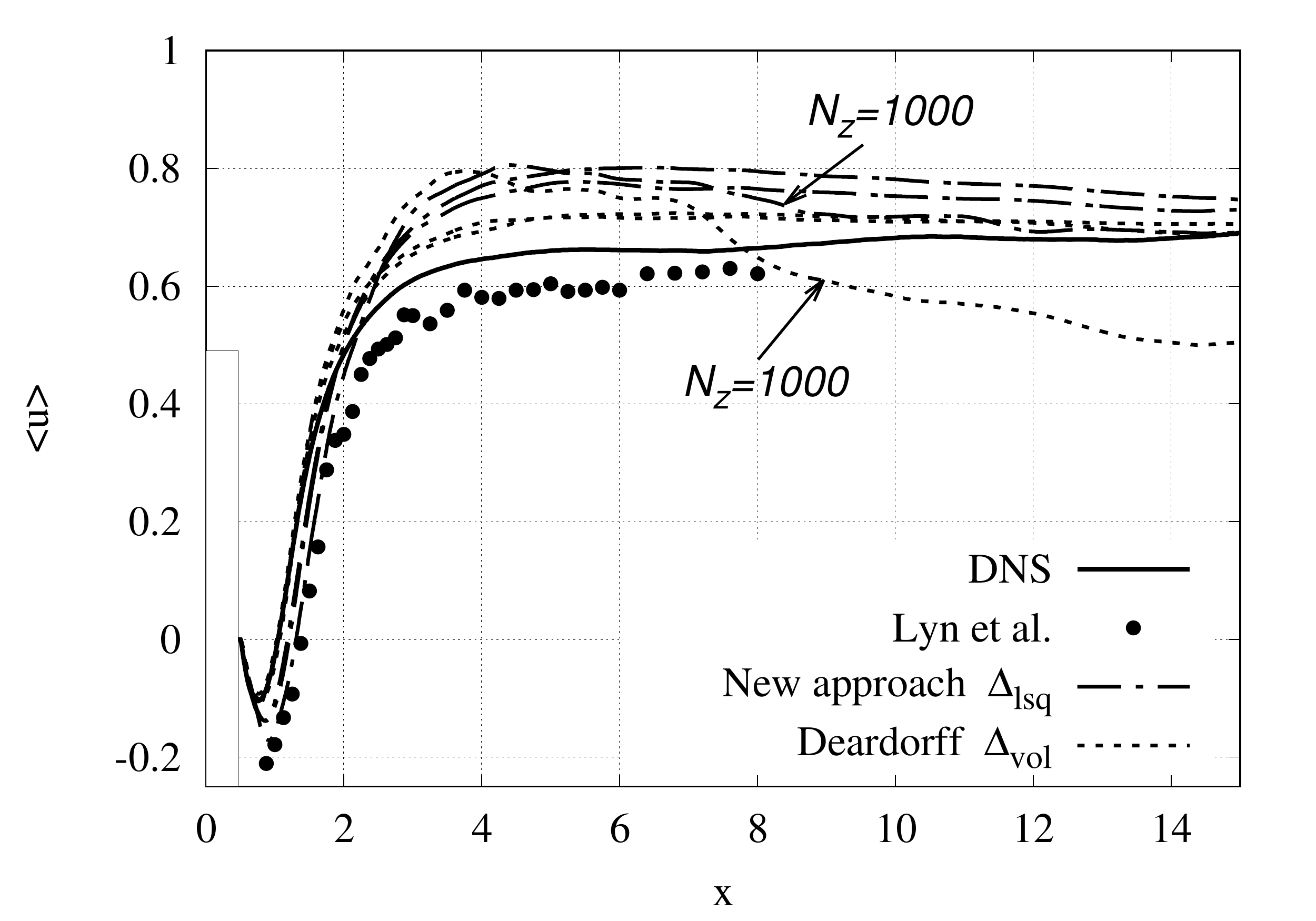}
	}
	\caption{The same as in Figure~\ref{CYL22K_U_BL}. In this case,
		average stream-wise velocity profiles are in the domain
		centerline. Experimental results of Lyn~\etal~\cite{LYN95} are also
		displayed for comparison.}
	\label{CYL22K_U_Wake2}
\end{figure}

\begin{figure}[!t]
	\centering{
		\includegraphics[angle=-90,width=0.63\textwidth]{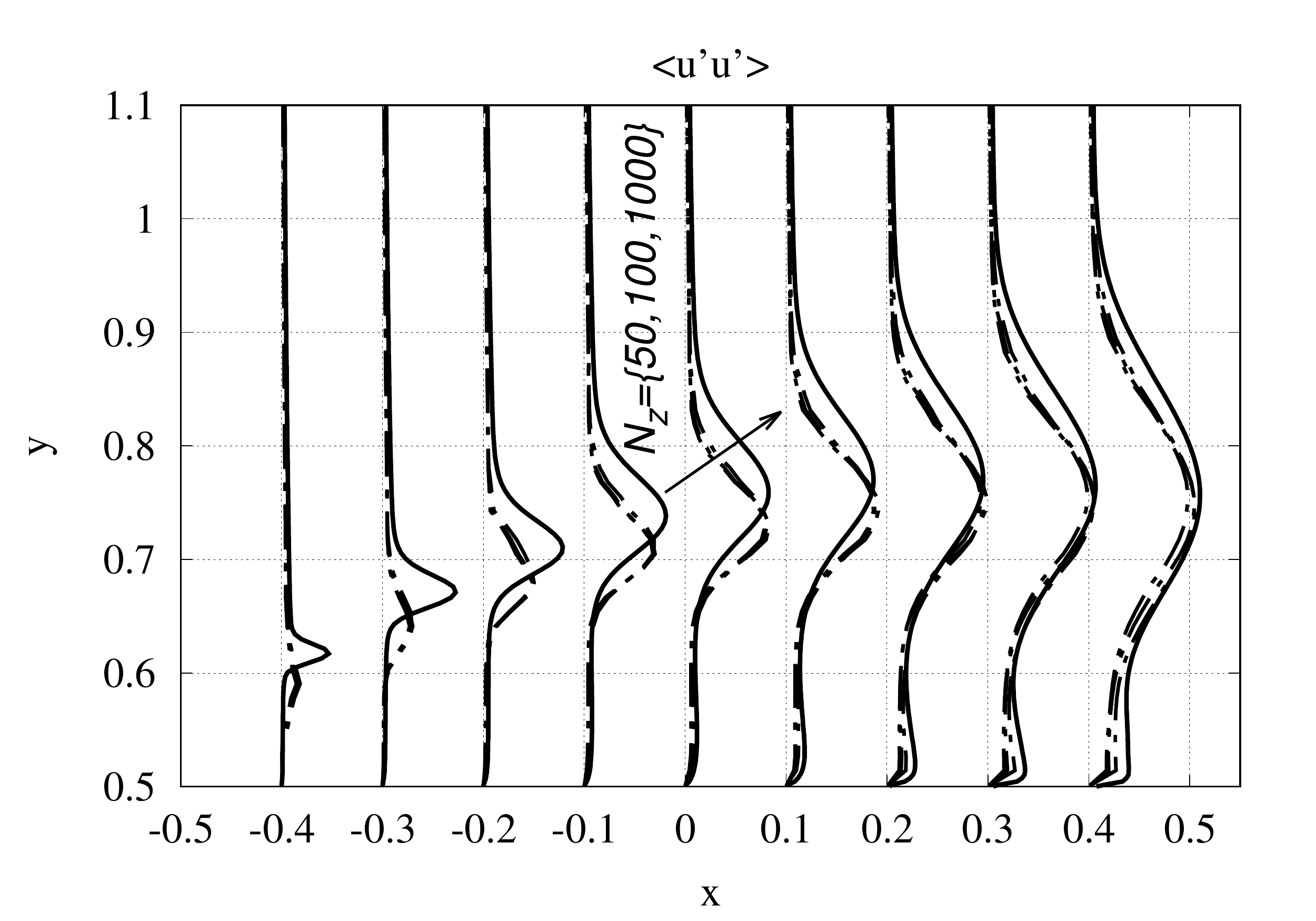}
		\includegraphics[angle=-90,width=0.63\textwidth]{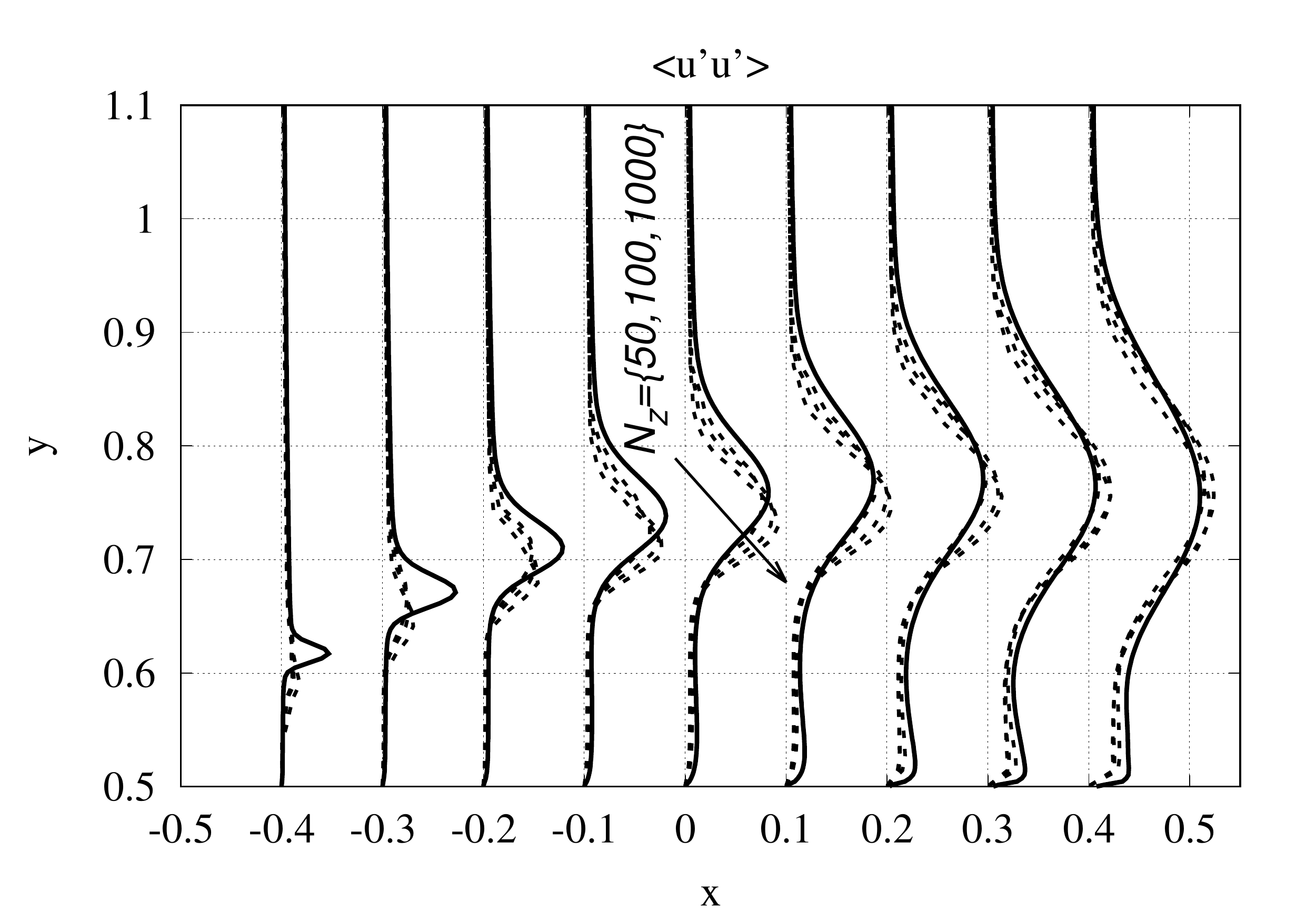}
	}
	\caption{{Profiles of the stream-wise Reynolds stresses,
			$\avgtime{\uvel^\prime \uvel^\prime}$, in the near-wall
			region. Results correspond to a turbulent flow around a square
			cylinder at $Re=22000$ carried out using the S3PQ
			model~\cite{TRI14-Rbased} on a set of unstructured meshes obtained
			from the extrusion in the span-wise direction, \ie
			~$\Nz=\{50,100,1000\}$, of the 2D mesh displayed in
			Figure~\ref{CYL22K_Mesh}. Solid lines correspond to the DNS
			results of Trias~\etal~\cite{TRI14-CYL}. Results obtained with the
			novel definition $\FLlsq$ (top) proposed in Eq.~(\ref{DeltaLsq})
			are compared with the classical definition proposed by Deardorff
			(bottom) given in Eq.~(\ref{DeltaDeardorff}).}}
	\label{CYL22K_UU_BL}
\end{figure}

\subsection{Flow around a square cylinder}

Finally, to test the performance of the proposed length scale with
unstructured meshes, the turbulent flow around a square cylinder has
been considered. In this case, the Reynolds number, $Re=UD/\nu=22000$,
is based on the inflow velocity, $U$, and the cylinder width,
$D$. This is a challenging test case for LES. {Apart} from the
well-known \VK vortex shedding in the wake region, this regime is
characterized by the clear presence of \KH vortical structures
produced by the flow separation at the leading edge of the
cylinder. The size of these vortices grows quickly, triggering
turbulence before they reach the downstream corner of the
cylinder. Actually, they break up into finer structures before being
engulfed into the much larger \VK vortices. An additional motivation
to choose this configuration is the fact that it has been studied
before in many numerical and experimental studies. The reader is
referred to our DNS study~\cite{TRI14-CYL} and references therein for
further details about the flow dynamics.

In {the current} work, we have carried out LES simulations on
unstructured meshes using the NOISEtte code for the simulation of
compressible turbulent flows in problems of aerodynamics and
aeroacoustics. It is based on the family of high-accuracy
finite-volume EBR (Edge-Based Reconstruction) schemes for unstructured
meshes~\cite{ABA16}. The EBR schemes provide at a low computing cost a
higher accuracy than most Godunov-type second-order schemes on
unstructured meshes. On translationally-invariant (structured) meshes
the EBR schemes coincide with high-order (up to sixth-order)
finite-difference schemes. Hybrid schemes with combination of upwind
and central-difference parts are used in LESs with
automatic adaptation of the weights of the components in order to
preserve numerical stability at a minimal numerical dissipation.

{For the simulations, the} 2D unstructured mesh displayed in
Figure~\ref{CYL22K_Mesh} has been extruded in the span-wise
direction. The resulting meshes have $19524 \times \Nz$ control
volumes, where $\Nz$ is the number of control volumes in the span-wise
direction. In this study we have considered three values for $\Nz$,
\ie~$\Nz=\{50,100,1000\}$. The first two meshes are reasonable for an
LES~\cite{MIN11} whereas the mesh with $\Nz=1000$ is
clearly too fine (even for a DNS~\cite{TRI14-CYL}). The 2D base mesh
{(see Figure~\ref{CYL22K_Mesh}, left)} is basically composed of
triangular elements, except for the region around the square cylinder
where there are skewed quadrilateral elements (see
Figure~\ref{CYL22K_Mesh}, right). The dimensions of the computational
domain are slightly smaller than in the DNS study~\cite{TRI14-CYL}:
$27D \times 27D \times 3D$ in the stream-wise, cross-stream and
span-wise directions, respectively. The upstream face of the cylinder
is located at $6.5D$ from the inflow and centered in the cross-stream
direction. The origin of coordinates is placed at the center of the
cylinder.

Similar to the turbulent channel flow, LES results
have been obtained with the S3PQ model~\cite{TRI14-Rbased} using two
definitions of $\Flength$: the new definition, $\FLlsq$, proposed in
Eq.~(\ref{DeltaLsq}) and the classical definition proposed by
Deardorff, given in Eq.~(\ref{DeltaDeardorff}). Results are compared
with the experimental data of Lyn~\etal~\cite{LYN95} and our
incompressible DNS results~\cite{TRI14-CYL} which are taken as a
reference. This DNS was carried out with a constant velocity profile,
$\vel = ( U, 0 ,0 )$, at the inflow, convective boundary conditions,
$\partial \vel / \partial t + U \partial \vel / \partial x = 0$, at
the outflow, Neumann boundary condition in the cross-stream direction,
$\partial \vel / \partial y = 0$, periodic boundary conditions in the
span-wise direction and a no-slip condition at the surface of the
cylinder. To make the comparison possible, present LESs are
carried out with analogous boundary conditions and at a nearly
incompressible Mach number ($M=0.1$). Results are presented in
dimensionless form where the reference length and velocity are {the cylinder width, }$D${,} and the inflow velocity,
$U$, respectively.

Results for the average eddy-viscosity, $\avgtime{\nut}$, divided by
the kinematic viscosity, $\nu$, are displayed in
Figures~\ref{CYL22K_nut} and~\ref{CYL22K_nut_zoom2}. Results for two
meshes, \ie~$\Nz=100$ (top) and $\Nz=1000$ (bottom), are shown. As
expected, values of $\nut$ obtained with Deardorff's length scale are
strongly affected by such abnormal mesh anisotropies. In this regard,
the ability of the new subgrid characteristic length, $\FLlsq$, to
adapt to these situations is remarkable. At first sight, the results
displayed in Figure~\ref{CYL22K_nut} look almost identical for
$\FLlsq$, whereas very significant differences are observed for the
Deardorff definition. A closer inspection (see
Figure~\ref{CYL22K_nut_zoom2}) reveals how both definitions of
$\Flength$ respond to the abrupt mesh transition between the near
obstacle region and the rest of the domain (see
Figure~\ref{CYL22K_Mesh}, right). The new subgrid characteristic
length tends to mitigate the effects of this mesh transition compared
with the results obtained with the Deardorff definition. This
difference becomes more evident for the mesh with $\Nz=1000$. 

{Sharp} discontinuities in $\nut$ may have severe negative
effects. Numerically, they can lead to more stringent time steps and
potentially cause instabilities. The former increases the
computational cost of the simulation, whereas the later can be solved
using a proper discretization of the viscous
term~\cite{TRI12-DmuS}. From a physical point of view, this abnormal
behavior of $\nut$ can negatively effect the quality of the results in
an uncontrolled manner. In this regard, results of the average
stream-wise velocity, $\avgtime{\uvel}$, in the near cylinder region
are displayed in Figure~\ref{CYL22K_U_BL}. In this case, both the
influence of the definition of $\Flength$ and the number of grid
points in the span-wise direction, $\Nz$, are rather small. This is a
consequence of the fact that the turbulence model itself has a
relatively low impact in the near-wall region, especially near the
upstream corner. Therefore, discrepancies with the DNS results in this
region can simply be attributed to insufficient grid resolution. This
region is actually characterized by the formation of small vortices in
the shear layer due to the \KH instability that are rapidly convected
downstream. The size of these vortices grows quickly, triggering
turbulence before they reach the downstream corner of the
cylinder. Interestingly, much better agreement with the DNS results
is achieved in this region. Apart from this, it is also interesting to
observe that results obtained with the new subgrid characteristic
length, $\FLlsq$, become closer to the DNS results when $\Nz$
increases. The results obtained with the Deardorff definition display
an opposite behavior. These trends remain the same in the wake region
where differences between both approaches are more visible. This is
clearly observed in the average stream-wise velocity profiles
displayed in Figure~\ref{CYL22K_U_Wake}. The anomalous behavior of the
Deardorff definition when refining in the span-wise direction becomes
more evident further downstream. The most relevant results in this
regard are the average stream-wise velocity profiles in the domain
centerline displayed in Figure~\ref{CYL22K_U_Wake2}. Results obtained
with $\Nz=1000$ and the Deardorff definition of $\Flength$ are
completely different from those obtained with $\Nz=50$ and $\Nz=100$,
showing that the definition of $\Flength$ itself can have a very
negative impact on the performance of an SGS model. Such an abnormal
behavior is not observed with the new subgrid characteristic length,
$\FLlsq$. 

{These results confirm the findings of the first two
  test cases (\ie~decaying isotropic turbulence and a turbulent
  channel flow): compared with the Deardorff definition, $\FLvol$, the
  proposed definition, $\FLlsq$, is much more robust with respect to
  mesh anisotropies. For those two cases, it was also seen that
  results using $\FLlsq$ are at least as good as the best results
  obtained with other definitions. Furthermore, it was also observed
  that these trends are even more evident for turbulent
  statistics. Here, the results for the stream-wise Reynolds stresses,
  $\avgtime{\uvel^\prime \uvel^\prime}$, in the near-wall region
  displayed in Figure~\ref{CYL22K_UU_BL} seem to confirm this. The robustness of the new definition, $\FLlsq$, in the
  shear layer region where there is almost a perfect match for the
  three meshes is remarkable ($\Nz=\{50,100,1000\}$), while differences are
  observed for the classical Deardorff definition, $\FLvol$. The new
  definition, $\FLlsq$, is even more robust if we consider that in
  this shear layer region there is an abrupt mesh transition from structured
  hexahedral elements to unstructured triangular prisms (see
  Figure~\ref{CYL22K_Mesh}, right).}

\section{Concluding remarks}

\label{conclusions}

In this work, a novel definition of the subgrid characteristic length,
$\Flength$, has been proposed with the aim to answer the following
research question: {\itshape Can we find a simple and robust definition of
  $\Flength$ that minimizes the effect of mesh anisotropies on the
  performance of SGS models?} In this respect, due to its simplicity
and mathematical properties we consider the flow-dependent $\FLlsq$
given in Eq.~(\ref{DeltaLsq}) a very good candidate. Namely, it is
locally defined, frame invariant, well bounded (see properties {\bfseries
  P1} and {\bfseries P2} in Section~\ref{properties}), and well conditioned, and
it has a low computational cost (property {\bfseries P5}). Moreover, a
simple extension of this length scale for unstructured grids (property
{\bfseries P4}) has been proposed in Section~\ref{unstr}: it basically
consists in replacing $\DelTen$ in Eq.~(\ref{DeltaLsq}) by the local
Jacobian, $\J_i$, defined in Eqs.~(\ref{locJacob})
and~(\ref{unstrJacob}). Finally, from the definition of $\FLlsq$ it is
obvious that it is dependent on the local flow topology given by the
gradient of the resolved velocity, $\G \equiv \nabla \F{\vel}$
(property {\bfseries P3}). In this respect, analytical analysis for simple
flow configurations points out the adequacy of the proposed
definition. Numerically, it has been successfully tested for
simulations of decaying homogeneous isotropic turbulence and a
turbulent channel flow at $Re_\tau = 395$ using (artificially) refined
grids. Comparisons with the classical length scale of Deardorff have
shown that the proposed definition is much more robust with respect to
mesh anisotropies. Due to these findings and its simplicity, we think
the currently proposed length scale has a great potential to be used
in subgrid-scale models in complex geometries where highly skewed
(unstructured) meshes are present.

\paragraph{Acknowledgments} This work has been financially supported by the \textit{Ministerio de
  Econom\'{i}a y Competitividad}, Spain (No.~ENE2014-60577-R). F.X.T. is
supported by a \textit{Ram\'{o}n y Cajal} postdoctoral contract
(RYC-2012-11996) financed by the \textit{Ministerio de Econom\'{i}a y
  Competitividad}, Spain. A.G. is supported by the Russian Science
Foundation (Project No.~15-11-30039). M.H.S. is supported by the Free
Competition in the Physical Sciences (Project No.~613.001.212),
which is financed by the Netherlands Organization for Scientific
Research (NWO). Part of this research was conducted at the CTR Summer
Program 2016: F.X.T., M.H.S. and R.W.C.P.V. thank the Center for
Turbulence Research (CTR) at Stanford University for hospitality and
financial support. 
This work has been carried out using computing resources of the Barcelona Supercomputing Center and the Federal collective usage center ``Complex for Simulation and Data Processing for Mega-science Facilities'' at NRC ``Kurchatov Institut'', \href{http://ckp.nrcki.ru/}{http://ckp.nrcki.ru/}.

\begingroup
\setlength\bibitemsep{0pt} 
\printbibliography
\endgroup

\end{document}